\documentclass[namedreferences,hyperref,optionalrh]{spr-sola}
\usepackage{graphicx}        
\usepackage{color}           
\usepackage{breakurl}        

\begin{document}

\begin{article}
\begin{opening}

\title{The Sun's Alfv\'{e}n Surface: Recent Insights and Prospects for the
\textit{\textbf{Polarimeter to Unify the Corona and Heliosphere}} (PUNCH)}

\author[addressref={affCUAPSLASP},corref,%
email={steven.cranmer@colorado.edu}]%
{\inits{S.R.}\fnm{Steven~R.}~\lnm{Cranmer}%
\orcid{0000-0002-3699-3134}}

\author[addressref={affUD,affGSFC}]%
{\inits{R.}\fnm{Rohit}~\lnm{Chhiber}%
\orcid{0000-0002-7174-6948}}

\author[addressref={affSwRI,affCUAPSLASP}]%
{\inits{C.R.}\fnm{Chris~R.}~\lnm{Gilly}%
\orcid{0000-0003-0021-9056}}

\author[addressref=affSydney]%
{\inits{I.H.}\fnm{Iver~H.}~\lnm{Cairns}%
\orcid{0000-0001-6978-9765}}

\author[addressref=affNRL]%
{\inits{R.C.}\fnm{Robin~C.}~\lnm{Colaninno}%
\orcid{0000-0002-3253-4205}}

\author[addressref=affPrince]%
{\inits{D.J.}\fnm{David~J.}~\lnm{McComas}%
\orcid{0000-0001-6160-1158}}

\author[addressref=affAPL]%
{\inits{N.E.}\fnm{Nour~E.}~\lnm{Raouafi}%
\orcid{0000-0003-2409-3742}}

\author[addressref={affUD,affGSFC}]%
{\inits{A.V.}\fnm{Arcadi~V.}~\lnm{Usmanov}%
\orcid{0000-0002-0209-152X}}

\author[addressref=affNCAR]%
{\inits{S.E.}\fnm{Sarah~E.}~\lnm{Gibson}%
\orcid{0000-0001-9831-2640}}

\author[addressref=affSwRI]%
{\inits{C.E.}\fnm{Craig~E.}~\lnm{DeForest}%
\orcid{0000-0002-7164-2786}}

\address[id=affCUAPSLASP]%
{Department of Astrophysical and Planetary Sciences,
Laboratory for Atmospheric and Space Physics,
University of Colorado, Boulder, CO, USA}

\address[id=affUD]%
{Department of Physics and Astronomy,
University of Delaware, Newark, DE, USA}

\address[id=affGSFC]%
{Heliophysics Science Division, NASA Goddard Space Flight Center,
Greenbelt, MD, USA}

\address[id=affSwRI]%
{Southwest Research Institute,
1050 Walnut Street, Suite 300, Boulder, CO, USA}

\address[id=affSydney]%
{School of Physics, University of Sydney, Sydney, NSW 2006, Australia}

\address[id=affNRL]%
{Space Science Division, Naval Research Laboratory,
Washington, DC 20375, USA}

\address[id=affPrince]%
{Department of Astrophysical Sciences, Princeton University,
Princeton, NJ 08544, USA}

\address[id=affAPL]%
{Johns Hopkins University Applied Physics Laboratory, Laurel, MD, USA}

\address[id=affNCAR]%
{National Center for Atmospheric Research,
3080 Center Green Drive, Boulder, CO, USA}

\runningauthor{S.\  R.\  Cranmer et al.}
\runningtitle{The Sun's Alfv\'{e}n Surface}

\begin{abstract}
The solar wind is the extension of the Sun's hot and ionized corona,
and it exists in a state of continuous expansion into interplanetary space.
The radial distance at which the wind's outflow speed exceeds the phase
speed of Alfv\'{e}nic and fast-mode magnetohydrodynamic (MHD) waves is
called the Alfv\'{e}n radius.
In one-dimensional models, this is a singular point beyond which most
fluctuations in the plasma and magnetic field cannot propagate back
down to the Sun.
In the multi-dimensional solar wind, this point can occur at different
distances along an irregularly shaped ``Alfv\'{e}n surface.''
In this article, we review the properties of this surface and discuss
its importance in models of solar-wind acceleration, angular-momentum
transport, MHD waves and turbulence, and the geometry of magnetically
closed coronal loops.
We also review the results of simulations and data analysis techniques
that aim to determine the location of the Alfv\'{e}n surface.
Combined with recent perihelia of \textit{Parker Solar Probe,} these
studies seem to indicate that the Alfv\'{e}n surface spends most of
its time at heliocentric distances between about 10 and 20 solar radii.
It is becoming apparent that this region of the heliosphere
is sufficiently turbulent that there often exist multiple (stochastic and
time-dependent) crossings of the Alfv\'{e}n surface along any radial ray.
Thus, in many contexts, it is more useful to make use of the concept of a
topologically complex ``Alfv\'{e}n zone'' rather than one closed surface.
This article also reviews how the \textit{Polarimeter to Unify the Corona
and Heliosphere} (PUNCH) mission will measure the properties of the
Alfv\'{e}n surface and provide key constraints on theories of
solar-wind acceleration.
\end{abstract}
\keywords{%
Heliospheric base;
Sun: corona;
Sun: magnetic field}
\end{opening}


\section{Introduction}
\label{sec:intro}

The solar corona is a region of the Sun's atmosphere defined by its
high temperature ($T > 10^{6}$~K), and the solar wind is defined
mainly by the presence of radially accelerating plasma that reaches
supersonic speeds within a few solar radii [$R_{\odot}$] of the
solar surface.
The corona is typically observed near the Sun with telescopes,
and the solar wind is typically probed further out with in-situ
particle and field detectors.
Despite these historical differences, these two regions overlap to
such an extent that it often makes no sense to specify a dividing
line between them.
Still, it is known that the strongest forces acting on the plasma
undergo a transition from being mostly magnetic, near the Sun, to
mostly hydrodynamic -- i.e.\  depending on gas-pressure gradients
and nonlinear inertial gas flow terms -- far from the Sun.
A common way to quantify this transition is to locate the
{\em Alfv\'{e}n surface,} the place where the radially increasing
solar-wind speed [$u$] exceeds the radially decreasing Alfv\'{e}n speed
[$V_{\rm A}$].
This article provides an overview of the various ways that the
Alfv\'{e}n surface is significant to our understanding of the physics
of the heliosphere, and it also discusses past, present, and future
attempts to measure its detailed properties.

The topic of this article has been called by a number of different names.
One-dimensional and time-steady models talk about an Alfv\'{e}n radius,
an Alfv\'{e}n point, or an Alfv\'{e}nic critical point.
With multi-dimensional models came the realization that there is no
single radial distance that has this property, so the community began
discussing the Alfv\'{e}n surface, or Alfv\'{e}nic critical boundary,
as a closed but nonspherical ``bubble.''
The synonymous term {\em heliobase} was also coined as a way to
discuss the Alfv\'{e}n surface's role as an inner boundary condition for
the larger heliosphere \citep{ZH10}.
We will tend to use the symbol $r_{\rm A}$ to refer to its heliocentric
radial distance, and the regions of the heliosphere with $r < r_{\rm A}$
and $r > r_{\rm A}$ are called sub-Alfv\'{e}nic and super-Alfv\'{e}nic,
respectively.
The vicinity of $r \approx r_{\rm A}$, which may contain multiple
crossings of the point at which $u = V_{\rm A}$,
is called the trans-Alfv\'{e}nic region or Alfv\'{e}n zone.

The present decade is an exciting time for studies of the Sun's
Alfv\'{e}n surface.
Most notably, \textit{Parker Solar Probe} \citep[PSP:][]{Fo16,Rao23}
has been the first spacecraft to cross it and measure the properties
of particles and fields in the sub-Alfv\'{e}nic region \citep{Ka21}.
However, an in-situ probe will never be able to measure the
global three-dimensional (3D) shape of the Alfv\'{e}n surface.
We anticipate the launch of the \textit{Polarimeter to Unify the
Corona and Heliosphere} (PUNCH), a NASA Small Explorer mission that will
perform regular, global, deep-field imaging of the accelerating
solar wind \citep{De22}.
One of the six main scientific questions to be addressed by the
PUNCH mission is: What are the evolving physical properties of the
Alfv\'{e}n surface?
PUNCH will produce the first global maps of the time-variable shape of
this key boundary region in the heliosphere, and
Working Group 1C of the PUNCH Science Team is in the process of
preparing to make the most of these measurements.
This article provides a snapshot of the current state of the science
and a summary of plans for this objective moving forward.

Section~\ref{sec:basics} of this article discusses several different
physical contexts in which the Alfv\'{e}n surface plays an important role.
Section~\ref{sec:locations} reviews theoretical and observational
attempts to measure $r_{\rm A}$ itself, and
Section~\ref{sec:punch} speculates about how the PUNCH mission will
improve on the existing measurements.
Lastly, Section~\ref{sec:conc} concludes by discussing broader
implications for other open questions in heliophysics and astrophysics.

\section{Definitions and Physical Processes}
\label{sec:basics}

In Section~\ref{sec:basics:crit}, we define the various critical points
found in solar wind models, including the Alfv\'{e}n radius.
Then we discuss the importance of the Alfv\'{e}n radius to some key 
phenomena in the corona and solar wind, such as
angular momentum transport (Section~\ref{sec:basics:angmom}),
the evolution of waves and turbulence (Section~\ref{sec:basics:turb}),
and the Sun's open/closed magnetic topology
(Section~\ref{sec:basics:loops}).

\subsection{Critical Points in a Time-Steady Wind}
\label{sec:basics:crit}

Here, we summarize the basics of how the solar wind is understood
to accelerate through several critical or singular points of the
magnetohydrodynamic (MHD) conservation equations.
These points are related to the linear propagation of information
through the system; i.e.\  they describe characteristics of
the underlying differential equations that govern the flow.
For other insightful reviews of these processes, see, for example,
\citet{Hu72}, \citet{Sk90}, \citet{LG97}, and \citet{Ow09}.

\citet{P58} solved the time-steady system described by the spherically
symmetric hydrodynamic outflow of an isothermal gas, with an
equation of motion given by
\begin{equation}
  \left[ u - \frac{c_i^2}{u} \right] \frac{{\rm d}u}{{\rm d}r} \, = \,
  \left[ \frac{2 c_i^2}{r} - \frac{GM_{\odot}}{r^2} \right] \,\,\, .
\end{equation}
This equation shows how the radial outflow speed $u(r)$ behaves as a
result of both gravity (with gravitational constant $G$ and solar mass
$M_{\odot}$) and the gas-pressure gradient force (here parameterized
by a constant sound speed $c_i$).
In this case, the flow is purely hydrodynamic (no magnetic forces),
and the radial acceleration ${\rm d}u/{\rm d}r$ remains finite when
both terms in square brackets are zero:
\begin{equation}
  u_{\rm crit} \, = \, c_i 
  \,\,\,\,\,\,\,\,\,
  \mbox{and}
  \,\,\,\,\,\,\,\,\,
  r_{\rm crit} \, = \, \frac{GM_{\odot}}{2 c_i^2} \,\,\, .
\end{equation}
This occurs at a ``critical radius'' that is equivalent to the
point where the flow transitions from subsonic ($u < c_i$) to
supersonic ($u > c_i$).
Note that a hotter corona exhibits a larger value of $c_i$ and a
smaller critical radius.
In that case, the outward gas-pressure gradient force exceeds the
inward force of gravity sooner.
MHD effects play no role in the time-steady momentum equation if the
magnetic field vector is oriented parallel to the flow-speed vector.
In the \citet{P58} model, this situation would imply a radially
oriented field.

Initially, Parker was criticized because the transonic critical-point
solution seemed unnaturally ``fine-tuned.''
Why should the Sun choose the one specific acceleration trajectory
that passes through both $r_{\rm crit}$ and $u_{\rm crit}$, rather
than one of the seemingly infinite other slower wind or ``breeze''
solutions where, say, $u < c_i$ everywhere?
However, when the actual inner and outer boundary conditions are
considered, Parker's transonic solution turns out to be something
akin to a stable attractor -- i.e.\  a solution toward which nonsteady
solutions tend to converge -- whereas the other solutions exhibit
instabilities \citep[see, e.g.][]{Ve94,Ve01,Ke20}.
In addition, the critical point loses some of its
importance when including {\em time-dependent} terms back into the
momentum conservation equation \citep[see, e.g.][]{Su82,HL97}.
In that case, the system tends to relax to stable transonic
acceleration without any regard for the supposed delicateness of
the critical solution.

The observational discovery of the supersonic solar wind
\citep{NS62} supported Parker's basic hydrodynamic approach.
In the decade that followed, the isothermal model was
generalized to allow for radial variations in temperature
\citep{P64}, as well as different temperatures for the
ions and electrons \citep{SH66} and directionally anisotropic
pressure tensors \citep{Ho70}.
\citet{WD67} extended this kind of model to two spatial dimensions;
i.e.\  the equatorial plane defined by the Sun's rotation axis
\citep[see also][]{BM76,Sk85,TC16}.
Since this scenario allows for the vector magnetic field
$\mbox{\boldmath $B$}$ and flow velocity $\mbox{\boldmath $u$}$
to no longer be aligned, it can account for the coexistence of forces
from both gas-pressure gradients and the magnetic field
(i.e.\  $\mbox{\boldmath $J$} \times \mbox{\boldmath $B$}$).
Thus, a generally accelerating outflow may pass through up to
{\em three} distinct critical points corresponding to the phase
speeds of obliquely propagating Alfv\'{e}n waves and fast/slow
magnetosonic waves.

\begin{figure}
\centerline{\includegraphics[width=0.99\textwidth,clip=]%
{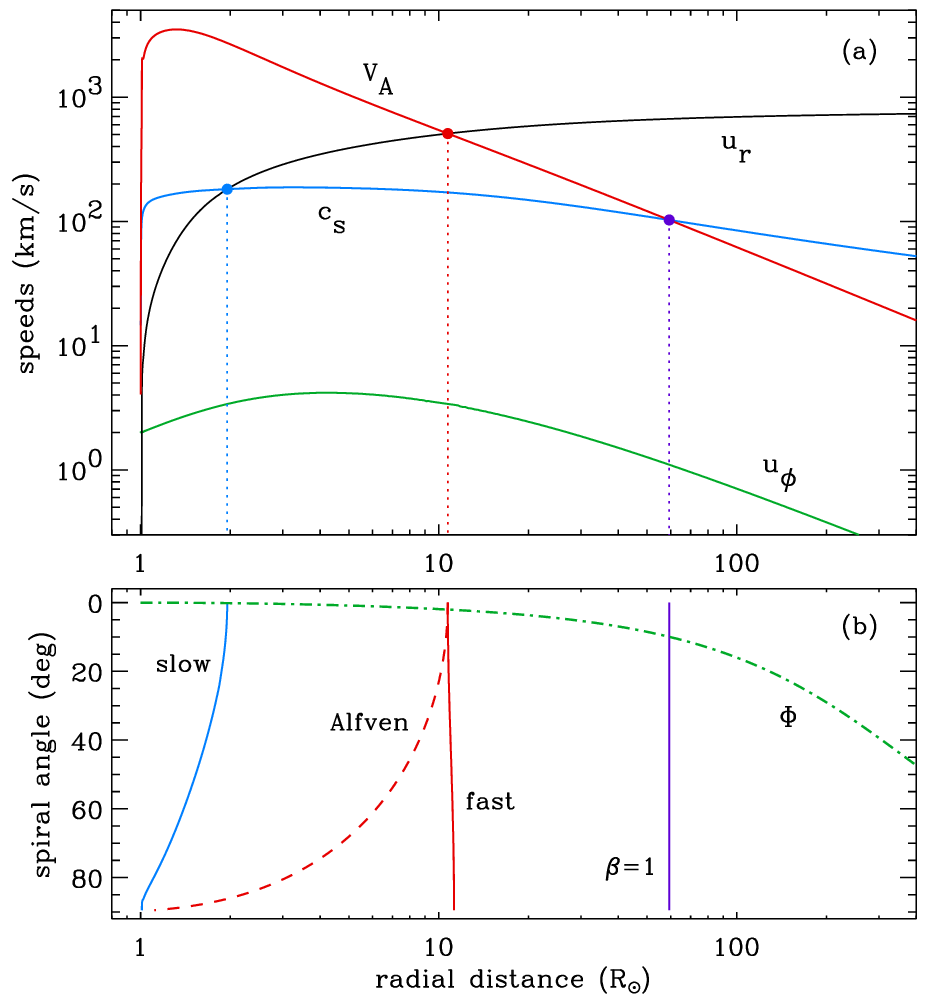}}
\caption{Example one-dimensional solar-wind model, with
quantities plotted vs.\  heliocentric radial
distance $r$, in units of solar radii $R_{\odot}$:
(\textbf{a}) radial outflow speed (\textit{black}),
sound speed (\textit{blue}), Alfv\'{e}n speed
$V_{{\rm A},r} = B_r/(4\pi\rho)^{1/2}$ (\textit{red}),
and azimuthal flow speed (\textit{green});
(\textbf{b}) locations of MHD critical points shown with respect
to both $r$ and the Parker spiral angle $\Phi$
(\textit{red and blue curves}),
plotted alongside the self-consistent $\Phi(r)$ for the
\citet{WD67} model (\textit{green dot-dashed curve}).
In both panels, \textit{purple} denotes
the $\beta = 1$ surface (see text).}
\label{fig01}
\end{figure}

Figure~\ref{fig01}a shows an example solar-wind model with
its three MHD critical points.
The radial acceleration is described by the polar coronal-hole
model of \citet{CvB07}, in which the corona is heated by
turbulent dissipation and the wind is also driven by
ponderomotive wave-pressure effects.
This model has a super-radially expanding magnetic field that is
similar to that found in other low-latitude parts of the corona;
e.g.\  equatorial coronal holes and quiet-Sun regions.
The one-fluid temperature (i.e.\  the average
of the proton and electron temperatures) is given by an
observationally constrained set of curves from multiple remote
and in-situ measurements in the high-speed solar wind
\citep{Cr20}.
Figure~\ref{fig01}a also shows the \citet{WD67} solution for
the azimuthal velocity in the ecliptic plane,
\begin{equation}
  u_{\phi}(r) \, = \, \Omega r \left[
  \frac{(M_{\rm A} r_{\rm A} / r)^2 - 1}{M_{\rm A}^2 - 1}
  \right] \,\, ,
  \label{eq:uphi}
\end{equation}
where the Alfv\'{e}nic Mach number $M_{\rm A} = u_r / V_{{\rm A},r}$
and the angular rotation rate at the Sun's surface is
$\Omega = 2.9 \times 10^{-6}$ rad~s$^{-1}$.
At low heights in the magnetically dominated corona, the rotation is
nearly rigid [$u_{\phi} \approx \Omega r$], which is related to
\citeauthor{F37}'s (\citeyear{F37}) law of iso-rotation.
At very large heights, ballistically flowing parcels of solar wind
approach a state of angular-momentum conservation
[$u_{\phi} \propto 1/r$].
At intermediate heights, there must be a transition between these
two disparate states.
Specifically, $M_{\rm A} = 1$ at the Alfv\'{e}n radius, so evaluating
$u_{\phi}$ requires L'H\^{o}pital's rule.
Making use of the fact that $\rho M_{\rm A}^2$ should remain constant
along a magnetic flux tube, and approximating the radial dependence
of the density as $\rho \propto r^{-n}$, one finds that
\begin{equation}
  u_{\phi} (r_{\rm A}) \, \approx \, \Omega r_{\rm A}
  \left( \frac{n-2}{n} \right) \,\,\, .
  \label{eq:uphicrit}
\end{equation}
For the model shown, $r_{\rm A} = 10.7 \, R_{\odot}$,
$n \approx 2.37$, and Equation~\ref{eq:uphicrit} correctly gives
$u_{\phi} \approx 3.3$~km~s$^{-1}$.
This is about an order of magnitude smaller than what it would have been
if the corona continued to rotate rigidly up to the Alfv\'{e}n radius.

It should be noted that different studies have used slightly different
definitions when determining the locations where $M_{\rm A} = 1$.
\citet{WD67} specifically used the ratio of the radial components
$u_r$ and $V_{{\rm A},r}$, with the latter computed from the radial
component of the magnetic field $B_r$.
\citet{By22} used $u_r$ and the magnitude of $B$,
\citet{Co15} and \citet{Ch22} used the magnitudes $u$ and $B$,
and \citet{KG00} used only the poloidal ($r$ and $\theta$) components
of both quantities.
However, the differences between these choices tend to be quite
insignificant because both the solar-wind flow and the magnetic field
are both mostly radial at the distances where $r_{\rm A}$ tends to occur.

Figure~\ref{fig01}b shows how the radial locations of the
MHD critical points vary as a function of the Parker spiral angle
$\Phi$ between $\mbox{\boldmath $u$}$ and $\mbox{\boldmath $B$}$.
For a purely radial flow and field ($\Phi = 0$), the slow-mode
magnetosonic critical point is just the classical transonic
critical point, and the Alfv\'{e}n and fast-mode critical
points coincide with one another at the point where $u_r = V_{\rm A}$.
However, the \citet{WD67} model provides a unique solution for
$\Phi$ as a function of radial distance, which is illustrated
in Figure~\ref{fig01}b and given by
\begin{equation}
  \tan \Phi \, = \, \frac{B_{\phi}}{B_r} \, = \,
  \frac{u_{\phi} - \Omega r \sin\theta}{u_r} \,\, .
\end{equation}
In the high-speed solar-wind model shown here, the field lines
are still mostly radial at the Alfv\'{e}n radius;
i.e.\  $|\Phi| \approx 2^{\circ}$ at $r_{\rm A}$,
and at that value the fast-mode critical point is only about
0.007~$R_{\odot}$ ahead of the Alfv\'{e}n point.

Figure~\ref{fig01}b also illustrates the location of the radial
distance at which the plasma $\beta$ ratio approximately equals unity.
Here, we apply the definition $\beta = c_s^2/V_{\rm A}^2$, which is
a version used frequently in collisionless plasma theory.
Defining the adiabatic sound speed as $c_s$, this expression for
$\beta$ differs by only a factor of 1.2 from the MHD definition
(the ratio of total gas pressure to magnetic pressure) when the
adiabatic exponent $\gamma = 5/3$.
For high-speed solar-wind models like the one shown here, the
$\beta = 1$ radius occurs well above the other critical points.
However, some slow-wind streams may experience additional crossings
of $\beta = 1$ near cusp-like null points at the tips of
helmet streamers (see Section~\ref{sec:basics:loops}).
\citet{De16} suggested that the heliospheric $\beta = 1$
radius is where large-scale turbulent plasma features undergo a
major transition in geometric aspect ratio.
Specifically, below this point they take the form of
magnetic-field-aligned rays or striations, and above this point they
become more isotropic cloud-like ``flocculations'' unconstrained
by the magnetic field.

There are a number of other factors that can affect the locations
and properties of the critical points.
As mentioned above, waves and turbulence exert a ponderomotive force
(``wave pressure'') that affects the location and speed of the
Parker critical point \citep{AC71,B71,J77,IH82}.
The existence of super-radial expansion of the magnetic-field lines
can give rise to the existence of multiple potential locations of
each critical point \citep{KH76}, and only a global examination of
the boundary conditions (i.e.\  what happens as $r \rightarrow \infty$)
determines which ones are actual critical points.
Although \citet{WD67} assumed a spherically expanding solar wind,
their expression for $u_{\phi}$ remains valid even with
super-radially expanding field lines.
Note, however, that the detailed form of Equation~\ref{eq:uphi}
changes when one takes account of additional
ion species -- e.g.\  $\alpha$ particles -- that flow at different
speeds than the background proton--electron plasma \citep{LL06}.

\subsection{Angular Momentum Transport}
\label{sec:basics:angmom}

The solar wind is primarily a source of mass loss for the Sun.
On average, its magnitude is $|{\rm d}M_{\odot}/{\rm d}t| = |\dot{M}|
\approx 10^9$~kg~s$^{-1}$, or about $3 \times 10^{-14}$
$M_{\odot}$~yr$^{-1}$ \citep[see, e.g.][]{Vi21}.
However, because the Sun is rotating, the solar wind is also
associated with a continuous loss of angular momentum, which causes
the Sun to spin down over time.
If the solar wind were released from the Sun's surface and immediately
began to conserve angular momentum (i.e.\  if $u_{\phi} \propto 1/r$),
the rate of solar angular momentum loss would be
\begin{equation}
  \frac{{\rm d}J}{{\rm d}t} \, = \,
  \frac{2}{3} \dot{M} \Omega R_{\odot}^2 \,\,\, ,
\end{equation}
where the factor of 2/3 assumes a spherically symmetric outflow
\citep[see][]{MP05}.
However, \citet{WD67} showed that a magnetized solar wind
increases the ``torque lever arm'' such that the total rate of
angular momentum loss is instead given by
\begin{equation}
  \frac{{\rm d}J}{{\rm d}t} \, = \,
  \frac{2}{3} \dot{M} \Omega r_{\rm A}^2 \,\,\, .
  \label{eq:WDdJdt}
\end{equation}
Thus, since $(r_{\rm A}/R_{\odot})^2 \approx 100$, this leads to
several orders of magnitude more rotational wind braking than was
suspected to exist originally.
Note that the appearance of $r_{\rm A}^2$ in Equation~\ref{eq:WDdJdt}
sometimes leads to the misconception that the plasma remains rigidly
rotating up to the Alfv\'{e}n radius.
The plot of $u_{\phi}$ in Figure~\ref{fig01}a indicates that
this is not the case.
In fact, most of the increased angular momentum transport in the
\citet{WD67} model is associated not with the rotating plasma, but
with Poynting stresses associated with $B_{\phi}$.

The present-day rate of rotational spindown for the Sun has been
estimated in various ways.
Astronomers have measured rotation frequencies $\Omega$ for
solar-type stars of a range of ages $t$, and there seems to exist a
power-law trend of $\Omega \propto t^{-1/2}$ that persists over
billions of years \citep{Sk72,Ba07}.
This relationship would be the natural result of
Equation~\ref{eq:WDdJdt} if $r_{\rm A} \propto B \propto \Omega$
and both $\dot{M}$ and the Sun's moment of inertia $I$ remained
constant over time; i.e.\  ${\rm d}\Omega / {\rm d}t \propto -\Omega^3$
\citep{DS72,Vi21}.
More direct attempts to measure ${\rm d}J/{\rm d}t$ from combinations of
in-situ plasma and magnetic-field data \citep{Pz83,Lj99,Fi19,Vs21b,Fi23}
have yielded results in adequate -- but still not precise -- agreement
with the basic theory.
\citet{Us18} used 3D simulations to model ${\rm d}J/{\rm d}t$ and
showed that the presence of turbulence can also affect the total loss of
angular momentum.
In addition, \citet{TC18} found that the near-Sun boundary conditions
used in simulations can have a strong impact on the resulting
behavior of angular-momentum loss in the heliosphere.

If the Sun is spinning down in a manner consistent with the
\citet{Sk72} scaling law [$\Omega \propto t^{-1/2}$], this would
imply a rate of period increase of roughly 0.02 seconds per century.
Such a change is currently unobservable, but it should be noted that
somewhat more rapid changes are observed for pulsars
\citep[e.g.][]{Ha13} and at least one main-sequence star \citep{Tw10}.
For the latter -- the magnetic B2-type star $\sigma$~Ori~E -- the
rate of period increase has been found to be approximately 8 seconds
per century, which appears to be consistent with the models of
rotational wind braking described above.
When considering other stars with a range of magnetic-field strengths,
wind mass-loss rates, and rotation rates, there arises an interesting
``zoo'' of dynamical phenomena that includes
shocked wind-fed disks, rigidly rotating stellar magnetospheres, and
stochastic competition between plasma infall and centrifugal breakout
\citep[see, e.g.][]{Ow09,uD17}.

\subsection{Importance to Waves and Turbulence}
\label{sec:basics:turb}

The Alfv\'{e}n radius is an important internal boundary condition
for models of MHD fluctuations in the solar wind.
For example, the simplest models of linear Alfv\'{e}n-wave evolution
along radial field lines often show a local maximum in the
wave velocity amplitude in the vicinity of $r_{\rm A}$.
Such models of WKB wave action conservation
(whose derivation does not require the use of the actual
Wentzel--Kramers--Brillouin approximation; see,
e.g.\  \citeauthor{BG68}, \citeyear{BG68};
\citeauthor{J77}, \citeyear{J77})
find that, in the absence of dissipation, the wave energy flux
$F = \rho v_{\perp}^2 V_{\rm A}$ should behave as
\begin{equation}
  F \, \propto \, \frac{B}{1 + M_{\rm A}^2} \,\, .
\end{equation}
Since the quantity $\rho M_{\rm A}^2$ should also be constant along
field lines, the transverse velocity amplitude of the waves
[$v_{\perp}$] is given by
\begin{equation}
  v_{\perp} \, \propto \, \rho^{-1/4} \left( 1 +
  \frac{\rho_{\rm A}}{\rho} \right)^{-1/2}
\end{equation}
where $\rho_{\rm A}$ is the plasma density at $r_{\rm A}$.
Thus, because $\rho$ is monotonically decreasing with increasing $r$,
$v_{\perp}$ increases with $r$ below the Alfv\'{e}n point
(i.e.\  $v_{\perp} \propto \rho^{-1/4}$ for $\rho \gg \rho_{\rm A}$).
The velocity amplitude reaches a local maximum exactly at the
Alfv\'{e}n point, and then it decreases above it, ultimately
scaling as $v_{\perp} \propto \rho^{+1/4}$ when $\rho \ll \rho_{\rm A}$.

Plasma properties at $r_{\rm A}$ are also required when computing
global models of non-WKB wave reflection \citep{HO80}.
In an inhomogeneous corona and heliosphere, some fraction of the
energy in outgoing waves may be reflected gradually and turned into
inward-propagating waves.
The Alfv\'{e}n radius is a singular point of the non-WKB transport
equations \citep[see also][]{Bk91,Ve93,MC94,Cr10}.
Specifically, the ratio of inward to outward velocity amplitudes
at $r = r_{\rm A}$, expressed in terms of inward [$Z_{-}$] and
outward [$Z_{+}$] \citet{E50} amplitudes, is found via another
application of L'H\^{o}pital's rule,
\begin{equation}
  \frac{Z_{-}}{Z_{+}} \, = \,
  \frac{| {\rm d}V_{\rm A}/{\rm d}r |}
  {\sqrt{\omega^2 + [({\rm d}u/{\rm d}r) -
  ({\rm d}V_{\rm A}/{\rm d}r)]^2}} \,\, ,
\end{equation}
where $\omega$ is the wave's angular frequency and the derivatives
are evaluated at $r_{\rm A}$.
For $r > r_{\rm A}$, the inward propagating Alfv\'{e}n waves cannot
travel all the way back down to the Sun.
However, they can still interact with the outward waves and undergo
nonlinear ``wave packet collisions'' that develop into MHD
turbulence and heat the plasma \citep[e.g.][]{Ir63,Kr65,Hs95}.
The analysis of non-WKB reflection in the presence of a strongly
turbulent cascade also requires treating the Alfv\'{e}n radius as a
singular point \citep{Dm02,CH09,CP19}.

The analysis of turbulent fluctuations measured in-situ can
be greatly complicated when the measurement probes approach the
Alfv\'{e}n radius.
Further out in the heliosphere (where $u \gg V_{\rm A}$), it is
common to see straightforward applications of 
\citeauthor{T38}'s (\citeyear{T38}) hypothesis of frozen-in
fluctuations.
This hypothesis assumes that waves and turbulent eddies are
essentially static in a reference frame that advects past the
spacecraft with the solar-wind velocity $\mbox{\boldmath $u$}$.
In other words, an observed frequency $\omega$ in the spacecraft
frame is assumed to be equivalent to
$\mbox{\boldmath $u$} \cdot \mbox{\boldmath $k$}$, where
$\mbox{\boldmath $k$}$
is a wavenumber intrinsic to the propagating fluctuation.
However, when a spacecraft approaches the vicinity of the Alfv\'{e}n
radius, the solar-wind velocity is no longer the dominant source of
motion.
Thus, modifications to Taylor's hypothesis need to be made
\citep[see, e.g.][]{Mt97,Kn15,BP18,Zn22}.
The study of heliospheric turbulence close to the Sun is still
an active field of research, and there have been theoretical predictions
of both depletions \citep{Ad19} and enhancements \citep{Ru20} of the
fluctuation amplitudes in the vicinity of the Alfv\'{e}n surface.

\subsection{Topological Boundary Between Open and Closed Fields}
\label{sec:basics:loops}

\begin{figure}
\centerline{\includegraphics[width=0.99\textwidth,clip=]%
{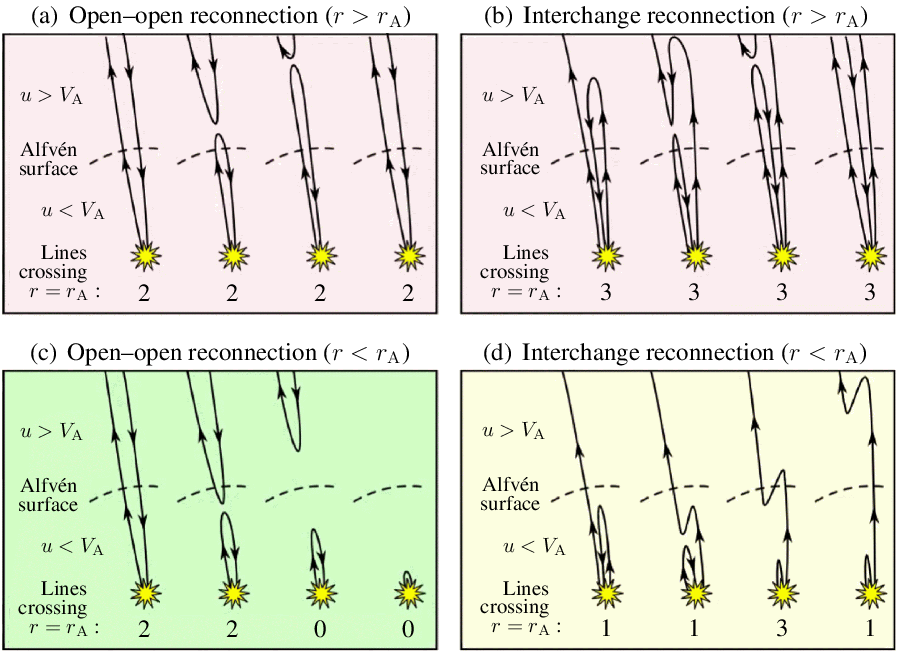}}
\caption{Temporal dependence of field lines crossing the
Alfv\'{e}n surface for various kinds of magnetic reconnection.}
\label{fig02}
\end{figure}

Once magnetic field lines extend above $r_{\rm A}$, they are unable
to propagate information along MHD characteristics back down to the Sun.
Thus, it is likely to be quite rare to find closed magnetic loops
that reach up to the Alfv\'{e}n radius or beyond.
Essentially, $r_{\rm A}$ can be considered to be a ``source surface''
of heliospheric magnetic flux \citep[e.g.][]{ZH10}.
It should be noted that, in the actual time-dependent heliosphere,
there are often closed field lines encountered at distances
far above $r_{\rm A}$ (i.e.\  detected with double-beamed electron strahls),
but these tend to be associated with coronal mass ejection (CME)
flux ropes, not parts of the ambient solar wind \citep{Go87,Sh00,Sm13}.
Nevertheless, if magnetic reconnection occurs in the outer corona,
it may be possible to trace the number of field lines that cross
the Alfv\'{e}n surface, over time, and this number will
vary in different ways depending on whether the reconnection
occurs above or below $r_{\rm A}$.
Figure \ref{fig02} shows a set of four possibilities, each
depending on the type of reconnection and where it occurs
\citep[see also][]{Ck02,Sw10,De12}.

Near the solar surface, the coronal magnetic field often obeys an
approximately force-free configuration
($\mbox{\boldmath $J$} \times \mbox{\boldmath $B$} \approx 0$)
that is satisfied by a potential field
($\mbox{\boldmath $J$} \propto \nabla \times \mbox{\boldmath $B$} = 0$,
so $\mbox{\boldmath $B$} = -\nabla \Psi$) in much of the volume
outside filaments and active regions.
The scalar potential $\Psi$ satisfies Laplace's equation, so it
can be expressed as a sum of multipolar spherical harmonics
with amplitudes specified by the photospheric lower boundary.
However, the acceleration of the solar wind produces MHD forces that
stretch open the field lines and cause them to depart from the
standard closed shapes (dipole, quadrupole, octupole, and so on).
This stretching is often modeled approximately using a radial
source-surface boundary condition at
$R_{\rm ss} \approx 2.5 \, R_{\odot}$ \citep{AN69,Sc69}
or the entire set of MHD equations is solved numerically
\citep[see, e.g.][]{PK71,Su99,Gm18,Us18}.

A consistent feature of both MHD simulations and observations
(from, e.g.\  the shapes of field lines inferred from visible-light
emission seen during total eclipses) is that
closed coronal loops do not typically extend all the way up
to the Alfv\'{e}n surface.
Although the largest ``helmet streamers'' can be seen to stretch
out to at least 10 to 15 $R_{\odot}$ (see Annie Maunder's historical
photo reproduced by \citeauthor{DF16} \citeyear{DF16}),
the magnetically closed loop-like features do not appear to extend
past radii of 3 to 4 $R_{\odot}$ \citep{Ri06,MY12}.
However, one may have expected to find the upper bound of the closed
magnetic field to be more like $r_{\rm A}$ itself, i.e.\  radial
distances of about 10 to 20 $R_{\odot}$ (see Section~\ref{sec:locations}).
This discrepancy has not been discussed extensively, although
\citet{Ow09} associated it with the prevalence of high-order multipole
components of the field that drop off with $r$ more rapidly than the
dominant dipole component.
Also, in some MHD models of streamers \citep[e.g.][]{KG00}, the
solar wind right above the cusp passes through an Alfv\'{e}n point
at a height even {\em lower} than Parker's sonic point.
The reduction in $V_{\rm A}$ near the cusp's null point gives the
solar wind's total pressure (gas and ram) the chance to substantially
distort the field lines, and even the plasma-$\beta$ ratio can
exceed unity in that region \citep{Lj98,Ll17}.

\begin{figure}
\centerline{\includegraphics[width=0.99\textwidth,clip=]%
{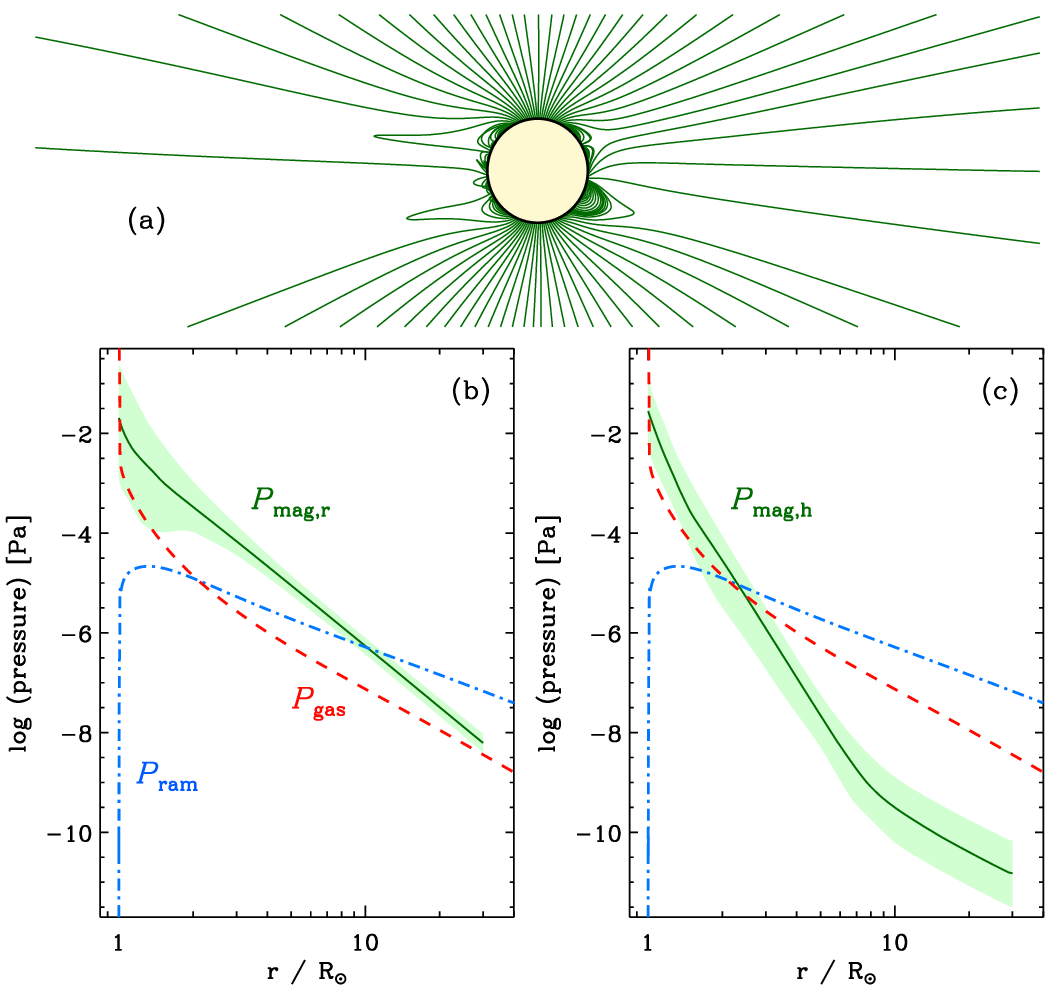}}
\caption{(\textbf{a}) Magnetic field lines traced from a time-steady
solution of the polytropic MHD conservation equations for
Carrington Rotation (CR) 2058.
(\textbf{b}) Comparison of radial dependences of time-steady gas
pressure (\textit{red dashed curve}) and ram pressure
($P_{\rm ram} = \rho u_r^2 / 2$, \textit{blue dot-dashed curve})
from Figure~\ref{fig01}, and radial magnetic pressure extracted
from the simulation (median: \textit{dark green solid curve,}
$\pm 1 \sigma$ bounds: \textit{light green region}).
(\textbf{c}) Same as panel (\textbf{b}), but with transverse
magnetic pressure.}
\label{fig03}
\end{figure}

Figure~\ref{fig03} illustrates these issues by highlighting the
difference between the magnetic pressure associated with $B_r$
and the magnetic pressure associated with the other (transverse
or horizontal) components of the field \citep[see also][]{Va03,RB17}.
For these plots, representative curves for the gas and ram pressure were
computed using the same example model shown in Figure~\ref{fig01}.
However, the details of the magnetic pressure come from numerical
models computed by the Magnetohydrodynamics Around a
Sphere (MAS) code \citep[see, e.g.][]{Ln99}
and made available by the MHDweb project
(\href{http://www.predsci.com/mhdweb/}%
{\textsf{www.predsci.com/mhdweb}}).
The photospheric boundary conditions were obtained from Carrington
Rotation (CR) 2058, from June--July 2007, with synoptic magnetogram
data from the \textit{Solar and Heliospheric Observatory} (SOHO)
\textit{Michelson Doppler Imager} \citep[MDI:][]{Sc95}.
The magnetic pressures in the corona were extracted at all latitudes
and longitudes, with only statistical medians and standard-deviation
ranges shown in the plots.

In the low-$\beta$ corona, flows parallel to the field act as
``beads on a string'' and do not feel significant Lorentz forces.
Thus, Figure~\ref{fig03}b shows that the height where the
ram pressure balances the radial magnetic pressure occurs at
$r_{\rm A} \approx 10 \, R_{\odot}$, as expected.
However, the shapes of the loops and streamers depend more on
the force balance perpendicular to $\mbox{\boldmath $B$}$, and
Figure~\ref{fig03}c shows that the associated transverse
magnetic pressure is overcome by both gas and ram pressure at
lower heights of order 2 to 3 $R_{\odot}$.
Thus, it is this range of heights that seems to be most naturally
associated with the cusps of the largest streamers.
As noted by \citet{Ow09}, the transverse components of the magnetic
field in the corona are dominated more by higher-order
(quadrupole and octupole) components of the global field, whose
strength drops off more rapidly than the dipole component.
Thus, the upper limit of closed-loop heights seems to be associated
with the same general processes that set the Alfv\'{e}n surface,
but this upper limit does not coincide with $r_{\rm A}$ itself.

\section{Locations of the Alfv\'{e}n Surface}
\label{sec:locations}

Having now summarized several ways that the Alfv\'{e}n surface is
important for our physical understanding of the corona and
solar wind, the next practical question to ask is:
Where is it?
Below we summarize quantitative estimates of $r_{\rm A}$ from
model-based predictions (Section~\ref{sec:locations:sims}),
in-situ measurements (Section~\ref{sec:locations:insitu}), and
coronagraphic imaging (Section~\ref{sec:locations:inflows}).

\subsection{Corona/Heliosphere Simulations}
\label{sec:locations:sims}

\begin{figure}
\centerline{\includegraphics[width=0.99\textwidth,clip=]%
{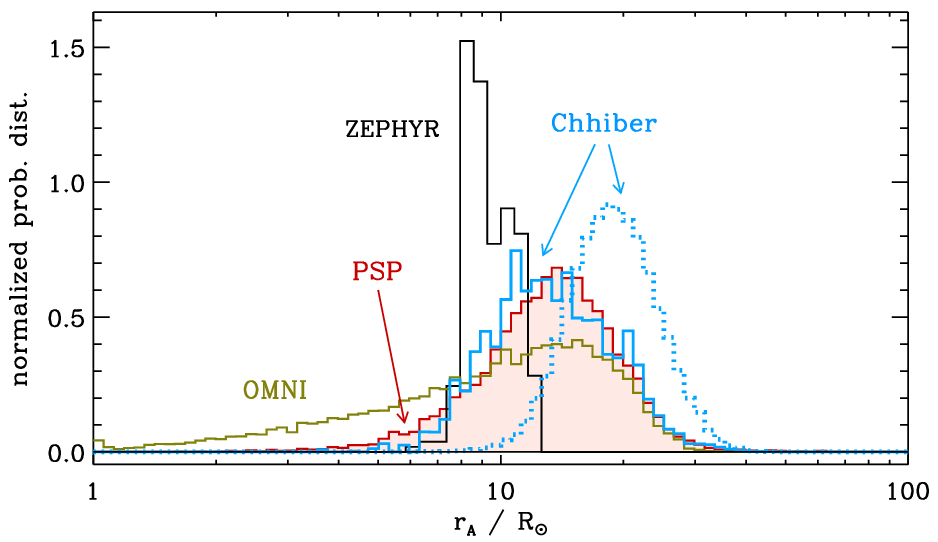}}
\caption{Comparison of independent estimates of Alfv\'{e}n radii,
expressed as probability-density histograms.
Model results include ZEPHYR (\textit{solid black curve}) and 3D
simulations from \citet{Ch22} (in-ecliptic: \textit{solid blue curve,}
all latitudes: \textit{dotted blue curve}).
Observationally derived radii include those from the OMNI
database at 1~AU (\textit{olive curve}) and from
PSP (\textit{red curve}).  See text for details.}
\label{fig04}
\end{figure}

The first numerical model that incorporated the Alfv\'{e}n surface was
that of \citet{WD67}, and their default set of in-ecliptic parameters
provided a value of $r_{\rm A} = 24.3 \, R_{\odot}$.
They also noted that the expected range of Alfv\'{e}n radii -- corresponding
to the observed variation of solar-wind properties at 1~AU -- would be
more like $r_{\rm A} \approx 15$ to 50~$R_{\odot}$.
Additional one-dimensional models of coronal heating and solar-wind
acceleration have been computed with the ZEPHYR code.
Figure~\ref{fig01} shows one specific model of this kind (for polar
magnetic field lines rooted in a coronal hole at solar minimum),
and Figure~\ref{fig04} shows a histogram of $r_{\rm A}$-values
constructed from 319 independent ZEPHYR models:
30 from the polar coronal hole, equatorial helmet streamer, and
active region models of \citet{CvB07}, and 289 from a high-resolution
study of open field lines connected to a diffuse quiet-Sun
region, by \citet{CvB13}.
For these models, $r_{\rm A}$ clusters rather tightly around a median
value of 9.16~$R_{\odot}$, with a standard deviation of only
1.24~$R_{\odot}$.

There have been a number of multi-dimensional global simulations that
track the variability of $r_{\rm A}$ as a function of longitude,
latitude, and solar-cycle activity
\citep{PK71,KG00,MP08,Co09,Pi11,Co15,Ch19}.
The shape of the Alfv\'{e}n surface follows the large-scale magnetic
polarity of the corona, often with local maxima in $r_{\rm A}$
occurring near the centers of large polar coronal holes, and local
minima occurring at the tops of helmet streamers.

\begin{figure}
\vspace*{0.03in}
\centerline{\includegraphics[width=0.99\textwidth,clip=]%
{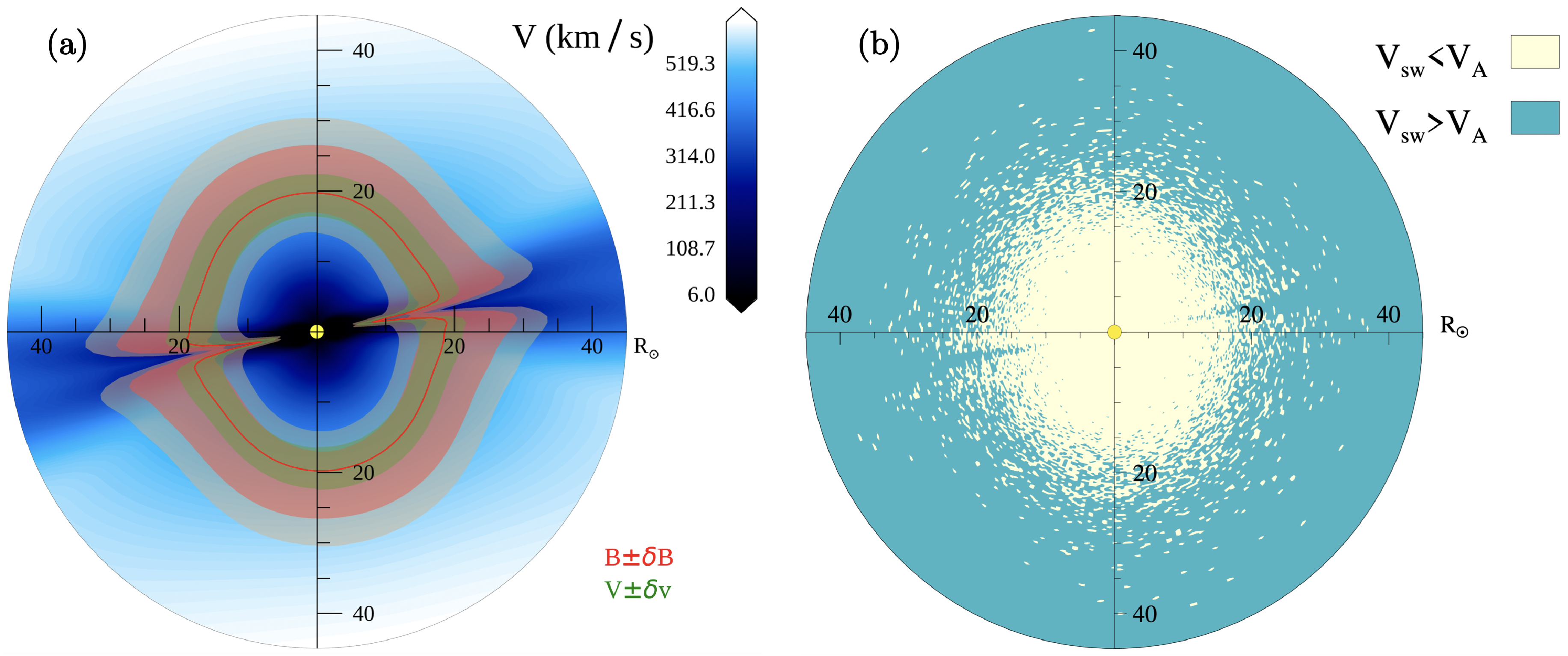}}
\vspace*{0.03in}
\caption{An example meridional plane cutting through one of the
3D simulations produced by \citet{Ch22}, showing:
(\textbf{a}) radial outflow speed (\textit{blue to white}),
overplotted with the mean shape of the Alfv\'{e}n surface
(\textit{red solid curve}) and ranges of locations over which the
Alfv\'{e}n surface occur over the simulation run when taking into
account magnetic fluctuations (\textit{light red region})
or velocity fluctuations (\textit{light green region}).
(\textbf{b}) Snapshot of the ``frothy'' separation between
sub-Alfv\'{e}nic (\textit{beige}) and super-Alfv\'{e}nic
(\textit{teal}) regions.}
\label{fig05}
\end{figure}

In order to better model the stochastic ``frothiness'' of the
Alfv\'{e}n zone as realistically as possible, \citet{Ch22} included
realizations of turbulent fluctuations in a 3D simulation.
In this model, there arise intermixed regions of sub-Alfv\'{e}nic and
super-Alfv\'{e}nic flow, with one radial ray passing through multiple
transitions where $u = V_{\rm A}$.
Histograms of values of $r_{\rm A}$ can be computed by taking
derivatives of cumulative distributions similar to the ones
shown in Figure~5 of \citet{Ch22}.
Figure~\ref{fig04} shows that when all latitudes and longitudes
are included, the distribution of Alfv\'{e}n radii has a substantially
higher median value (20.1~$R_{\odot}$) than do the ZEPHYR models.
However, when only simulation data for $\pm$4$^{\circ}$ of latitude
above and below the ecliptic plane were chosen -- specifically for a
model that reproduces CR~2215 -- the median value of $r_{\rm A}$
is lower (15.1~$R_{\odot}$) and in better agreement with the
observationally inferred values discussed in more detail below.
Figure~\ref{fig05} illustrates these simulations, and
\citet{Ch22} discuss further how these patchy structures may be
important for driving the largest energy-containing scales of MHD
turbulence in the solar wind.

\subsection{In Situ Extrapolations and Detections}
\label{sec:locations:insitu}

There have been quite a few studies of in-situ
spacecraft data that involved extrapolating radial trends
from traditionally observed distances ($r > 0.3$~AU, prior to PSP)
inwards to the vicinity of the Alfv\'{e}n radius.
Each of these studies tends to make slightly different assumptions
about the radial trends of $u$ and $V_{\rm A}$ -- and also different
assumptions about which conservation laws are used in the
extrapolation process -- but they often seem to converge on values
for $r_{\rm A}$ between about 10 and 40~$R_{\odot}$
\citep[see, e.g.][]{MR84,EM00,Kt10,Go14,TC16,TC18,KK19,Liu21,Vs21a}.
More recently, combinations of in-situ and remote-sensing data
have been used to put additional constraints on the most likely
values of $r_{\rm A}$ \citep{Wx21,Td21}, and these methods often
point to a lower range of about 8 to 20~$R_{\odot}$.

Perhaps the most straightforward of these extrapolation methods combines
the assumption of a constant solar-wind speed, mass-flux conservation
far from the Sun ($\rho \propto r^{-2}$) and magnetic-flux conservation
far from the Sun ($B_r \propto r^{-2}$) to find that one then expects
the Alfv\'{e}nic Mach number to increase with radial distance as
$M_{\rm A} \propto r$.
Thus, measuring $M_{\rm A} \gg 1$ in interplanetary space provides an
anchor-point to extrapolate linearly back to the point where this
quantity is equal to unity.
\citet{Cr21} showed how this method can be modified to account for
the small amount of radial acceleration in the solar-wind speed
between $r_{\rm A}$ and the in-situ measurement point.

Figure~\ref{fig04} shows our attempt to extrapolate to $r_{\rm A}$
using 11 years of in-ecliptic OMNI
(\href{https://omniweb.gsfc.nasa.gov/}%
{\textsf{omniweb.gsfc.nasa.gov}})
data taken at 1~AU \citep{KP05} from 2008 to 2018.
The extrapolation method is discussed in more detail by \citet{Cr21}.
CMEs and gaps in the OMNI database were removed, the former using
criteria given by \citet{XB15}, and mean values of of the solar-wind
parameters were taken using two-hour bins.
Note that \citet{Cr21} gave results for only the high-speed solar wind, but
the OMNI histogram shown in Figure~\ref{fig04} accounts for all speeds.
The resulting distribution of values of $r_{\rm A}$ is highly skewed,
with a modal maximum at 15.4~$R_{\odot}$ and a median at 10.9~$R_{\odot}$.

\begin{figure}
\centerline{\includegraphics[width=0.98\textwidth,clip=]%
{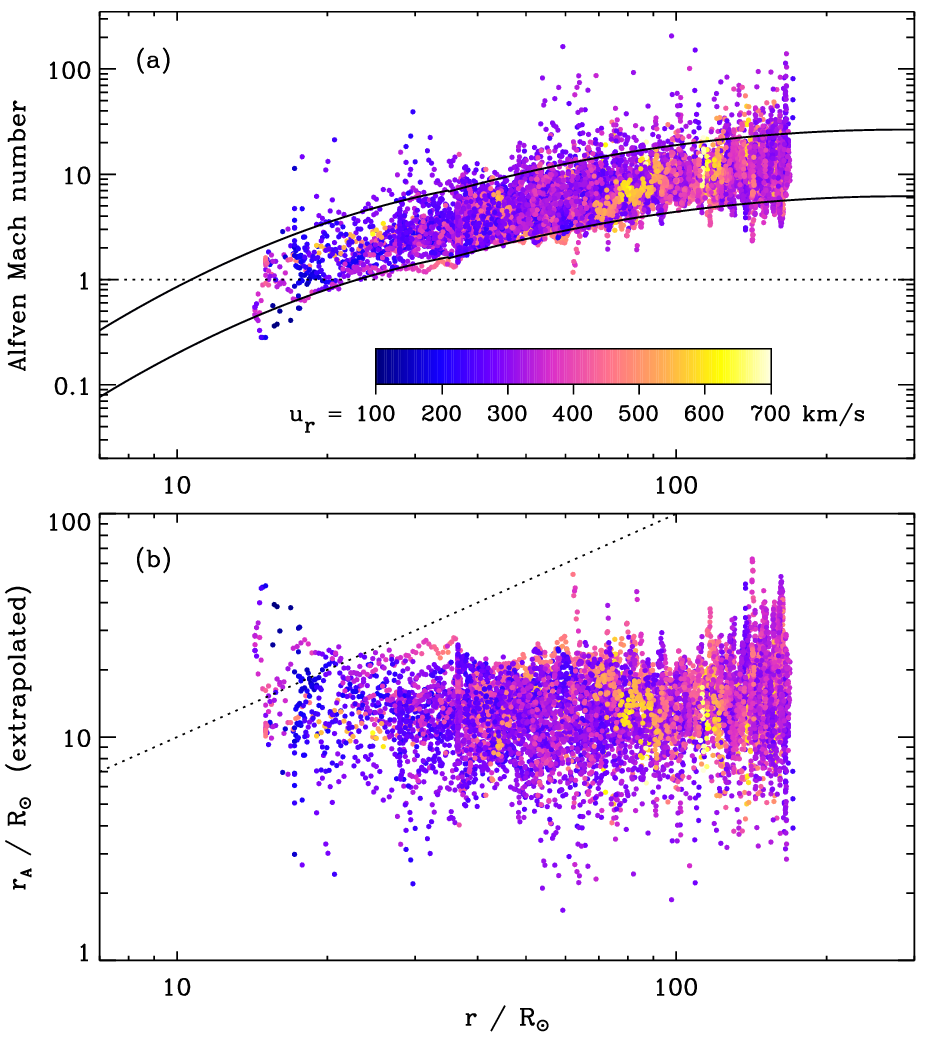}}
\caption{Preliminary analysis of PSP data over its first 13 perihelia
(October 2018 to October 2022).
(\textbf{a}) Alfv\'{e}nic Mach number $M_{\rm A}$ versus heliocentric
distance, with measured \textit{solar-wind speeds shown as symbol color.}
\textit{The solid curves} are provided only to indicate the approximate
lower and upper bounds of the data.
(\textbf{b}) Extrapolated locations of $r_{\rm A}$, plotted versus
the heliocentric distance of PSP when each measurement was made.
In both panels, \textit{the dotted line} indicates local crossings of
the Alfv\'{e}n surface.}
\label{fig06}
\end{figure}

Of course, now that the PSP spacecraft has repeatedly crossed into
the sub-Alfv\'{e}nic heliosphere
\citep[see, e.g.][]{Ka21,By22,Zg22,Zh22,Liu23},
the extrapolation techniques described above can be tested and validated,
at least statistically.
Figure~\ref{fig06}a shows preliminary PSP data for the Alfv\'{e}nic Mach
number $M_{\rm A}$ versus heliocentric distance.
We computed this quantity as the ratio of $u_r$ to the total Alfv\'{e}n
speed $V_{\rm A}$ (evaluated with the magnitude $B$).
To obtain these parameters, we began with one-hour-averaged merged
data, produced as a part of NASA's Coordinated Heliospheric
Observations (COHO), that combines Level-3 proton data from the
\textit{Solar Probe Cup} \citep[SPC:][]{Case20}, a component of the
\textit{Solar Wind Electrons Alphas and Protons} (SWEAP) suite \citep{Ka16},
with Level-2 fluxgate magnetometer data from the FIELDS suite \citep{Ba16}.
Because these data were not always available for the most recent
PSP perihelia, we followed \citet{By22} by also incorporating
additional proton radial velocity and density data from
the SWEAP \textit{Solar Probe Analyzer--Ion} (SPAN-I), with
data rejected if any of the SPAN-I quality-flag bits
0, 2, 3, 8, 10, or 11 were set.
Also, we incorporated additional plasma density data (with
$n_p = n_e/1.1$) from quasi-thermal noise measurements with the FIELDS
\textit{Radio Frequency Spectrometer} \citep[RFS:][]{Pu17}
and additional Level-2 fluxgate magnetometer data.
Data provided at higher resolution were binned
into one-hour averages and used only when the the COHO merged data
were unavailable.

We applied the OMNI extrapolation method discussed above to our
PSP dataset.
An extra step in the analysis was added to account for measurements not
at 1~AU, but the same basic algorithm from \citet{Cr21} remained in use.
Figure~\ref{fig06}b plots these extrapolated values of $r_{\rm A}$
versus the radial distance of PSP at the time of each measurement.
Note that there is no dominant trend in these derived values as a
function of heliocentric distance.
The median value for the resulting distribution of $r_{\rm A}$ values
was 13.4~$R_{\odot}$, with a standard deviation of 5.4~$R_{\odot}$.
Figure~\ref{fig04} shows the corresponding histogram, and its
resemblance to that corresponding to the near-ecliptic simulation of
\citet{Ch22} is striking.
The PSP histogram also shares a few properties with the histogram
computed from OMNI data at 1~AU.
For example, the two median values fall within $\pm 1$ standard
deviation of one another, and the upper edges of both distributions
are nearly identical.
Although it is too early to know for sure, it is likely that data
from PSP will end up providing better estimates of $r_{\rm A}$ than
data taken at 1~AU due to that spacecraft's more frequent proximity
to the Alfv\'{e}n zone.

Lastly, it is important to note that using in-situ data to
extrapolate down to the sub-Alfv\'{e}nic corona provides benefits
beyond just locating the Alfv\'{e}n surface.
Data-driven maps of the vector velocity and magnetic field
in the corona can constrain models of turbulence, corotating
interaction regions, and suprathermal particle scattering
\citep{TC18,TC19}.
Mapping field lines down to the solar surface can help identify the
origin sites of different types of fast and slow solar-wind streams
\citep[e.g.][]{Lu02,Lw04}.
The measurement of cross-field (possibly flux-tube-like) structure
in the in-situ data has also been used to speculate about
the survival of granular or supergranular scales from the Sun's
surface out to 1~AU \citep{Bv08,TC18,Ba21}.
These techniques, together with those that map from the corona out
to interplanetary space \citep[e.g.][]{RL11,Rs19}, are useful
in improving forecasts of the time-variable solar wind.

\subsection{Observations of Coronal Inflows}
\label{sec:locations:inflows}

\begin{figure}
\centerline{\includegraphics[width=0.99\textwidth,clip=]%
{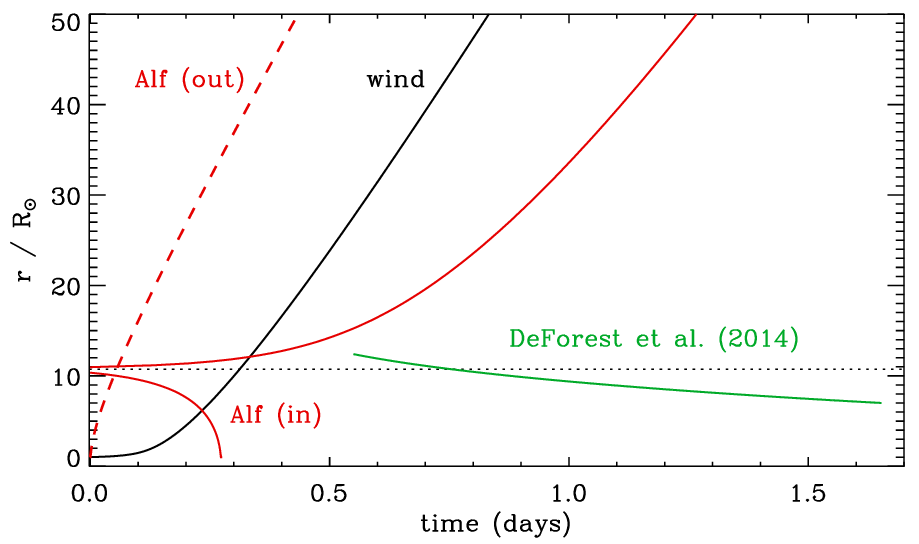}}
\caption{Height versus time plots for radial flows that accelerate with
the solar wind (\textit{black solid curve}), with an outward-propagating
Alfv\'{e}n wave (\textit{red dashed curve}), or with an inward-propagating
Alfv\'{e}n wave (\textit{red solid curves}).
Model parameters are the same as in Figure~\ref{fig01}, with
$r_{\rm A} = 10.7 \, R_{\odot}$ (\textit{black dotted line}).
Also shown is a fit to the inward-propagation ``ridge'' observed
in a polar coronal hole by \citeauthor{De14}
(\citeyear{De14}; \textit{green solid curve}).
For all curves, absolute start times are arbitrary;
i.e.\  any of them can be shifted to the left or right by any amount.}
\label{fig07}
\end{figure}

Coronagraphs and other off-limb imaging instruments have been used for
decades to measure outwardly propagating intensity fluctuations that
are believed to act as passive tracers (``leaves in the wind'') and
thus provide data on the radial acceleration of the solar wind
\citep[see, e.g.][]{Sh97,Ab16}.
However, there have been rarer detections of plasma flows that
go from higher to lower radii over time, and these can be used to
put limits on the location of the Alfv\'{e}n surface.
Figure \ref{fig07} shows how a feature propagating radially -- here using
the one-dimensional model of Figure~\ref{fig01} -- would move in radius
versus time, depending on whether it moves with the unperturbed
wind speed [$u_r$] or on linear Alfv\'{e}nic characteristics
[$u_r \pm V_{\rm A}$].
Note that true inward propagation seems to be possible only for the
``minus'' Alfv\'{e}n characteristic, for $r < r_{\rm A}$.

There have been quite a few different types of inflows observed near
the Sun.
Near the solar surface, there exist supra-arcade downflows
in active regions \citep[e.g.][]{Sv12} as well as ``coronal rain'' that
propagates down as dense clumps along coronal loops \citep{An15,Ms19},
though many of these flows probably do not connect to the solar wind.
At larger heights of about 2 to 6~$R_{\odot}$, there are frequent
blob-like downflows seen in the vicinity of streamer cusps
\citep{SW14,SD17,Ly20}.
The most distant inflows were found by \citet{De14}, using data from the
COR2 coronagraph on the \textit{Solar Terrestrial Relations Observatory}
\citep[STEREO-A:][]{How08}.
The detection of the weak signals associated with these inflows required
the use of customized background subtraction and filtering based on
combined spatial and temporal Fourier transforms.

Specifically, for observations made at solar minimum in 2007,
\citet{De14} found inbound features out to at least 12~$R_{\odot}$
in polar coronal-hole regions, and out to at least
15~$R_{\odot}$ in the streamer belt.
Figure~\ref{fig07} shows an integrated version of the coronal-hole
``inflow ridge'' measured by \citet{De14}, which was reported originally
in terms of radial velocity versus height.
These measurements indicated not only substantial deceleration -- from
speeds of about 80 km~s$^{-1}$ at 12.4~$R_{\odot}$ to only about
20 km~s$^{-1}$ at 7~$R_{\odot}$ -- but also a reduction in magnitude
of the parcel's acceleration that leads to a characteristic
``concavity'' in the velocity-versus-height diagram.
These data appear to confirm that the Alfv\'{e}n radius
must be at a height greater than 12~$R_{\odot}$ for this component of
the solar wind.

However, \citet{Te16} noted that the properties of the observed
inflow ridge do not match the expected behavior of inwardly
propagating Alfv\'{e}n waves.
The latter should accelerate when getting closer to the Sun, not
decelerate.
\citet{Te16} also found that other types of linear MHD waves
(i.e.\  obliquely propagating fast or slow magnetosonic waves) may
decelerate as they approach the Sun, but they do not
agree with the concavity of the measured inflow ridge.
One type of model that was found to match the data is one in which
a parcel of plasma undergoes a snowplow-like mass enhancement
over time, which leads to substantial deceleration as it is
entrained into the background flow.
\citet{Cr21} produced additional models of snowplow mass-gain
in conjunction with hydrodynamic drag forces between the parcel and
the ambient solar wind.
This kind of model could be consistent with bursty exhausts from
coronal sites of magnetic reconnection.
\citet{Cr21} also noted that the initial downflow speeds from nonlinear
features (e.g.\  shocks, jets, or shear instabilities) could be
somewhat supra-Alfv\'{e}nic, so they may be able to begin at slightly
larger distances than $r_{\rm A}$ and still propagate down towards
the Sun.

Lastly, it should be noted that a large amount of theoretical speculation
has been performed on behalf of a relatively small quantity of existing
imaging data in the vicinity of the Alfv\'{e}n surface.
Many more additional examples of inflowing parcel kinematics need to be
measured at other locations and times in the solar cycle.
Also, it should be noted that Alfv\'{e}nic fluctuations, by themselves,
do not tend to produce the density fluctuations that are most readily
observable as variations in off-limb Thomson-scattered intensity.
Thus, future theoretical models need to include a more self-consistent
description of either linear or nonlinear compressible features
that propagate at speeds similar to $V_{\rm A}$ and thus can help
probe the location of the Alfv\'{e}n surface.

\section{Discussion: Prospects for PUNCH}
\label{sec:punch}

The PUNCH mission will image the outer corona and inner heliosphere
at radial distances between 6 and 180 $R_{\odot}$, with a temporal
cadence ranging from four minutes (at $r < 80 \, R_{\odot}$) to
35 minutes (for the entire field of view).
To measure the flow speeds and accelerations of features in the
solar wind, the PUNCH team will use proven methods of
spatio-temporal Fourier filtering \citep{De14} and a number of
flow-tracking algorithms \citep[e.g.][]{CV06,CS08,AI15}.
This is being done in collaboration with
Working Group 1A of the PUNCH Science Team, whose goal is to map out
the spatial and temporal evolution of the ``young'' solar wind
\citep[see, e.g.][]{At23}.
PUNCH has been designed with a sensitivity and resolution to enable
the accumulation of a huge database of inbound-feature measurements,
and these will provide unprecedented information about the variable
location of the Alfv\'{e}n surface.

Although PUNCH will use linear polarization to perform 
three-dimensional localization of large structures like CMEs
\citep{De17}, it is not yet known to what extent this extra
information can be used to improve the tracking of smaller features
within the ambient solar wind.
Still, it may be the case that synergy between PUNCH and other future
missions that include off-limb polarimetry -- e.g.\  the
\textit{Association of Spacecraft for Polarimetric and Imaging
Investigation of the Corona of the Sun} \citep[ASPIICS:][]{Ga18}, or
the \textit{Coronal Diagnostic Experiment} \citep[CODEX:][]{Nw20} -- may
provide new perspectives on the use of linear polarization for
measuring the properties of the global solar wind.

As mentioned above, PUNCH measures off-limb density inhomogeneities, so
it does not directly see incompressible Alfv\'{e}n waves.
However, both slow-mode and fast-mode magnetosonic waves exhibit
density oscillations.
If, for example, PUNCH observes fast-mode waves propagating radially,
these could be used to probe the location of the Alfv\'{e}n surface
because they flow at speeds of roughly $\pm V_{\rm A}$ relative to the
solar wind \citep[see also][]{Te16}.
The extended corona may also contain nonlinearly
steepened features such as shocks, jets, reconnection exhausts,
or the end-products of shear instabilities.
These can produce observable density perturbations that flow at speeds
of order $V_{\rm A}$ as well.

The plane-of-sky speeds provided by PUNCH flow-tracking
will need to be analyzed in order to estimate the
true vector velocities of these features in 3D space.
This will be aided by {\em forward modeling} of these kinds of
features \citep[see, e.g.][]{Gi16,Gr20,GC21,MF22},
together with the development of robust uncertainty bounds on
this kind of 3D localization.
The properties of the measured density inhomogeneities will also need
to be correlated with the ambient properties of the solar wind and
the global magnetic geometry of the corona.
Specifically, PUNCH data will provide many examples of both:
i) the direct measurement of $V_{\rm A}$ at $r_{\rm A}$, from
knowing the bulk solar-wind speed there, and
ii) maps of the plane-of-sky shape of the Alfv\'{e}n surface.
These can help discriminate between competing solar-wind models and
provide large-scale context for future multi-messenger campaigns with
PSP and Solar Orbiter \citep[see, e.g.][]{MP23}.

It will be interesting to learn whether or not all measured inflow
tracks will have the same speed and acceleration trends (i.e.\  concavity)
as the one set of events studied by \citet{De14} and \citet{Te16}.
It is possible that the one inbound feature measured by COR2 was
the ``tip of the iceberg'' and thus is not representative of the
broader population of lower-contrast features to be seen by PUNCH.
Because these other features may have weaker density fluctuations,
they may not undergo substantial amounts of snowplow-like mass-gain.
Thus, they may have speeds more representative of linear MHD waves.
We expect to explore trends in the inflow dynamics as a function of
the relative intensity enhancements (i.e.\  relative density
enhancements) of blobs, which will help validate the physics included
in models that include mass-gain and drag forces \citep[e.g.][]{Cr21}.

\begin{figure}
\centerline{\includegraphics[width=0.97\textwidth,clip=]%
{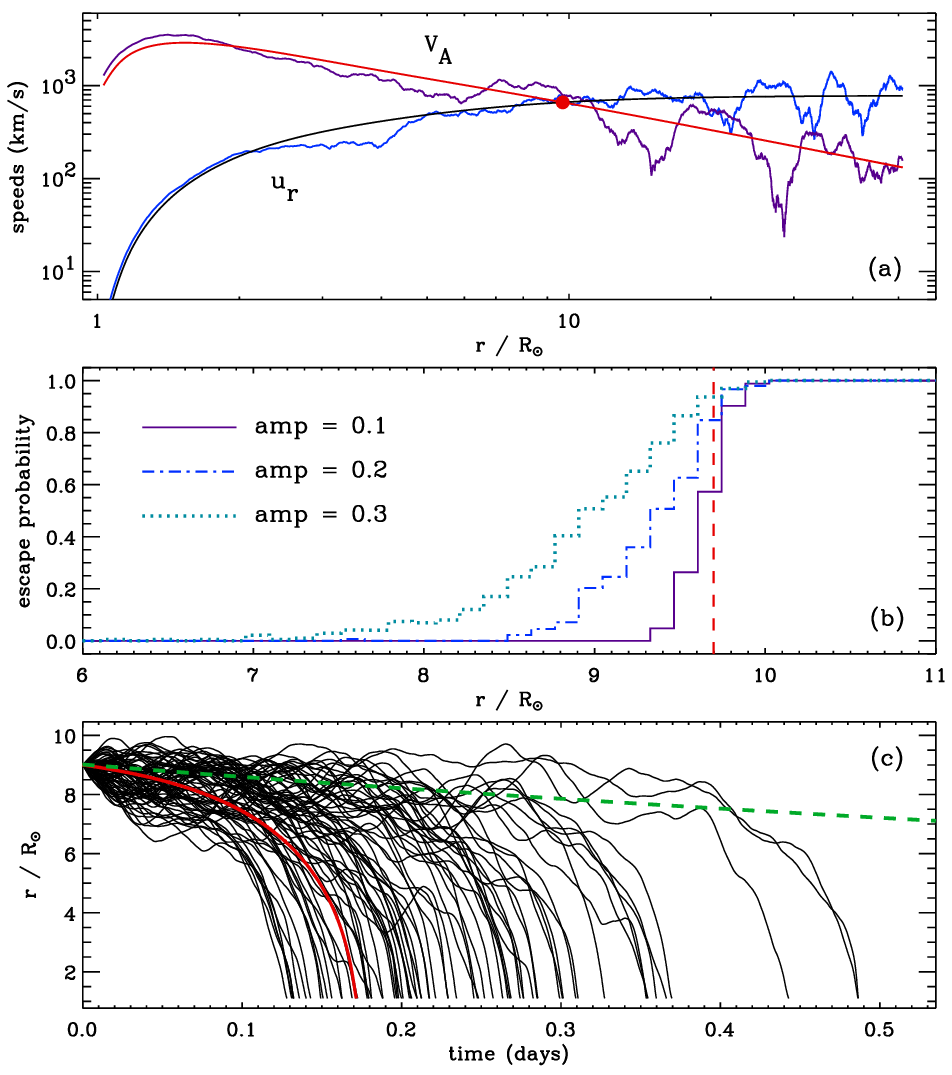}}
\caption{Preliminary model of stochastic motions in the solar wind.
(\textbf{a}) Example snapshot of $u_r$ and $V_{\rm A}$, both with no
turbulence (\textit{black and red curves}) and with turbulence included
at a fractional amplitude of 0.3 times the time-steady values
(\textit{blue and violet curves}).
(\textbf{b}) Escape probability of a parcel that flows with
local radial speed $(u_r - V_{\rm A})$, averaged over 10,000
trials for each value of turbulence amplitude.  The time-steady
value of $r_{\rm A}$ is shown with a \textit{dashed red line.}
(\textbf{c}) Downward trajectories that include turbulence at an amplitude
of 0.3 (\textit{black curves}) and one that does not (\textit{red curve}),
compared with observed COR2 inflow data (\textit{green dashed curve}).}
\label{fig08}
\end{figure}

It is also very likely that the region around the
Alfv\'{e}n radius is highly turbulent, so there may exist multiple
crossings in a frothy ``Alfv\'{e}n zone'' \citep[see also][]{De18,Ch22}.
One implication of this, which PUNCH may be very well-suited to
observe, is that individual blobs would undergo stochastic
(random-walk-type) deflections when near the Alfv\'{e}n zone.
Thus, they would propagate alternately in and out for a while before
heading decidedly either towards the Sun or the outer heliosphere.
Figure~\ref{fig08} shows the result of a simple one-dimensional
simulation that accounts for this stochasticity.
A continuous Kolmogorov-type spectrum of fluctuations
(i.e.\  power proportional to $f^{-5/3}$) was sampled at 100 discrete
frequencies $f$ sampled between $10^{-4}$ and 1 Hz, with random phases.
Then, one independently constructed time-series was imposed on the
background wind speed $u_r(r)$, and another was imposed on the
Alfv\'{e}n speed [$V_{\rm A}$], each with the same renormalized
relative amplitude, as shown in Figure~\ref{fig08}a.

Because models of non-WKB wave reflection usually show a preponderance
for outward waves in the corona, our simulated fluctuations were
assumed to propagate along outward Alfv\'{e}n characteristics
(i.e.\  at $u_r + V_{\rm A}$) over time.
However, we then chose to focus on the temporal evolution of parcels that
attempt to approach the Sun on inward characteristics [$u_r - V_{\rm A}$].
Due to the in/out motion of multiple Alfv\'{e}n radii in this model, 
it becomes possible for some parcels that begin at $r < r_{\rm A}$ to
eventually escape, and for some parcels that begin at $r > r_{\rm A}$ to
eventually flow downwards and reach the Sun.
Figure~\ref{fig08}b illustrates this by plotting histograms of a
statistical ``escape probability'' constructed from three sets of
10,000 random trials.
The three sets had different relative amplitudes for the imposed
fluctuations, and one can see that as this amplitude approaches zero,
the escape probability approaches a step function that is zero for
$r < r_{\rm A}$ and one for $r > r_{\rm A}$.
PSP observations indicate that these relative amplitudes are
reasonable for the radii shown here \citep{Ch22solo}.

Lastly, Figure~\ref{fig08}c shows a collection of inflow trajectories
for parcels that started at $r = 9 \, R_{\odot}$ (slightly below the
time-steady Alfv\'{e}n radius, which for this model was 9.7~$R_{\odot}$)
and eventually flowed down to the Sun.
A parcel flowing inwards along unperturbed inward Alfv\'{e}n
characteristics (i.e.\  with no ``froth'') would reach the Sun in only
about 0.17~days, but the distribution of perturbed parcels has a
longer mean flow time.
These perturbed parcels become trapped in the trans-Alfv\'{e}nic zone
due to being pivoted in the in/out direction multiple times, and
this effectively {\em decelerates} them.
Thus, this effect may help explain the observed inflow trend from
\citet{De14}.
We anticipate that PUNCH will have the sensitivity, cadence, and
spatial resolution to be able to detect these kinds of alternating
inflows and outflows for parcels buffeted by turbulence.
In fact, if PUNCH can measure the lifetimes, sizes, and relative
densities of these parcels, it can put new constraints on models
of MHD turbulence in the solar wind.

\section{Conclusions}
\label{sec:conc}

The objective of this article has been to review our present-day
understanding of the Sun's Alfv\'{e}n surface and to look forward
to how the PUNCH mission will improve that understanding.
This is the goal of Working Group 1C of the PUNCH Science Team,
and it connects to the higher-level goal of the mission's
Science Objective 1 (``to understand how coronal structures become
the ambient solar wind'').
Earlier, we discussed how the flow-tracking activities of
Working Group 1A are necessary to provide the measurements of
time-dependent flow speeds that can help map the Alfv\'{e}n surface.
In addition, what we learn about the frothy and stochastic
Alfv\'{e}n zone from PUNCH will be also be highly
relevant to Working Group 1B, whose goal is to determine how
micro-structures and turbulence form and evolve in the solar wind
\citep{Vn21,Pc23}.

Understanding the Alfv\'{e}n surface associated with a magnetized wind
is also relevant for solar systems beyond our own.
For example, it is likely that six out of the seven planets orbiting
the nearby M-dwarf TRAPPIST-1 spend most of their time inside
the star's sub-Alfv\'{e}nic zone \citep{Gf17}.
This tends to create planetary magnetospheres quite different in
character from those seen in our solar system.
For example, the motion of close-in exoplanets may disturb the plasma
sufficiently to induce strong energy fluxes that go back down to
the star \citep[e.g.][]{Saur13,Mats15}, producing visible chromospheric
starspots \citep{Shk08} or bursts of radio emission \citep{PV23}.
There are fascinating opportunities for synergy between studies of
our own Alfv\'{e}n surface, which we can study up close and in detail,
and those of other stars, which allow us to sample a much broader range
of plasma parameters \citep{GS23}.


\begin{acknowledgments}
The authors thank Sean Matt, Alex Chasapis, Riddhi Bandyopadhyay,
and Dusan Odstrcil for valuable discussions.
We are also grateful to the anonymous referee, guest editor, and
editor-in-chief, who made many constructive suggestions that have
improved this article.
The authors acknowledge NASA's Space Physics Data Facility
and COHOWeb for access to the OMNI and PSP data.
The National Center for Atmospheric Research is a major facility
sponsored by the National Science Foundation under Cooperative
Agreement 1852977.
\textit{Parker Solar Probe} was designed, built, and is now operated
by the Johns Hopkins Applied Physics Laboratory as part of NASA's Living
with a Star (LWS) program (contract NNN06AA01C).
Support from the LWS management and technical team has played a
critical role in the success of the \textit{Parker Solar Probe} mission.
The authors also thank the FIELDS team (PI: Stuart Bale, UC Berkeley) and
the SWEAP team (PI: Justin Kasper, BWX Technologies) for providing
data to the archive.
This research made extensive use of NASA's Astrophysics Data System (ADS).
\end{acknowledgments}

\begin{fundinginformation}
This work was supported by the National Aeronautics and Space
Administration (NASA) via contract 80GSFC18C0014 for the PUNCH
Small Explorer mission.
Additional support came from NASA via grants 80NSSC20K1319,
80NSSC18K1648, and 80NSSC22K1020,
and from the National Science Foundation (NSF) via grant 1613207.
\end{fundinginformation}

\begin{ethics}
\begin{conflict}
The authors declare that they have no conflicts of interest.
\end{conflict}
\end{ethics}

\end{article} 

\begin{thebibliography}{}

\bibitem[\protect\citeauthoryear{{Abbo et al.}}{2016}]{Ab16}
Abbo, L., Ofman, L., Antiochos, S.K., Hansteen, V.H., Harra, L.,
Ko, Y.-K., Lapenta, G., Li, B., Riley, P., Strachan, L., von Steiger, R.,
Wang, Y.-M.: 2016,
Slow solar wind: Observations and modeling.
\textit{Space Sci.\  Rev.} \textbf{201}, 55.
\href{https://doi.org/10.1007/s11214-016-0264-1}{DOI}.
\href{https://ui.adsabs.harvard.edu/abs/2016SSRv..201...55A/abstract}{ADS}.

\bibitem[\protect\citeauthoryear{{Adhikari et al.}}{2019}]{Ad19}
Adhikari, L., Zank, G.P., Zhao, L.-L.: 2019,
Does turbulence turn off at the Alfv\'{e}n critical surface?
\textit{Astrophys.\  J.} \textbf{876}, 26.
\href{https://doi.org/10.3847/1538-4357/ab141c}{DOI}.
\href{https://ui.adsabs.harvard.edu/abs/2019ApJ...876...26A/abstract}{ADS}.

\bibitem[\protect\citeauthoryear{{Alazraki and Couturier}}{1971}]{AC71}
Alazraki, G., Couturier, P.: 1971,
Solar wind acceleration caused by the gradient of Alfv\'{e}n wave pressure.
\textit{Astron.\  Astrophys.} \textbf{13}, 380.
\href{https://ui.adsabs.harvard.edu/abs/1971A&A....13..380A/abstract}{ADS}.

\bibitem[\protect\citeauthoryear{{Altschuler and Newkirk}}{1969}]{AN69}
Altschuler, M.D., Newkirk, G.: 1969,
Magnetic fields and the structure of the solar corona, I: Methods of
calculating coronal fields.
\textit{Solar Phys.} \textbf{9}, 131.
\href{https://doi.org/10.1007/BF00145734}{DOI}.
\href{https://ui.adsabs.harvard.edu/abs/1969SoPh....9..131A/abstract}{ADS}.

\bibitem[\protect\citeauthoryear{{Antolin et al.}}{2015}]{An15}
Antolin, P., Vissers, G., Pereira, T.M.D., Rouppe van der Voort, L.,
Scullion, E.: 2015,
The multithermal and multi-stranded nature of coronal rain.
\textit{Astrophys.\  J.} \textbf{806}, 81.
\href{https://doi.org/10.1088/0004-637X/806/1/81}{DOI}.
\href{https://ui.adsabs.harvard.edu/abs/2015ApJ...806...81A/abstract}{ADS}.

\bibitem[\protect\citeauthoryear{{Atti\'{e} and Innes}}{2015}]{AI15}
Atti\'{e}, R., Innes, D.E.: 2015,
Magnetic balltracking: Tracking the photospheric magnetic flux.
\textit{Astron.\  Astrophys.} \textbf{574}, A106.
\href{https://doi.org/10.1051/0004-6361/201424552}{DOI}.
\href{https://ui.adsabs.harvard.edu/abs/2015A&A...574A.106A/abstract}{ADS}.

\bibitem[\protect\citeauthoryear{{Atti\'{e} et al.}}{2023}]{At23}
Atti\'{e}, R., Thompson, B.J., Moraes Filho, V., Tremblay, B.,
Viall, N.M., Gallardo-Lacourt, B., Provornikova, E., Malanushenko, A.: 2023,
Mapping solar wind flows with PUNCH.
\textit{Solar Phys.,} in preparation.

\bibitem[\protect\citeauthoryear{{Bale et al.}}{2016}]{Ba16}
Bale, S.D., Goetz, K., Harvey, P.R., Turin, P., Bonnell, J.W.,
Dudok de Wit, T., Ergun, R.E., MacDowall, R.J., Pulupa, M.,
Andre, M., Bolton, M.,
Bougeret, J.-L., Bowen, T.A., Burgess, D., Cattell, C.A.,
Chandran, B.D.G., Chaston, C.C., Chen, C.H.K.,
Choi, M.K., Cranmer, S.R., Diaz-Aguado, M., Donakowski, W.,
Drake, J.F., Farrell, W.M., Fergeau, P., Fermin, J.,
Fischer, J., Fox, N., Glaser, D., Goldstein, M., Gordon, D.,
Hanson, E., Harris, S.E., Hayes, L.M., Hinze, J.J., Hollweg, J.V.,
Horbury, T.S., Howard, R., Hoxie, V., Jannet, G., Karlsson, M.,
Kasper, J.C., Kellogg, P.J., Kien, M.,
Klimchuk, J.A., Krasnoselskikh, V.V., Krucker, S., Lynch, J.J.,
Maksimovic, M., Malaspina, D.M., Marker, S., Martin, P.,
Martinez-Oliveros, J., McCauley, J., McComas, D.J.,
McDonald, T., Meyer-Vernet, N., Moncuquet, M., Monson, S.J.,
Mozer, F.S., Murphy, S.D.,  Odom, J., Oliverson, R., Olson, J.,
Parker, E.N., Pankow, D., Phan, T., Quataert, E.,
Quinn, T., Ruplin, S.W., Salem, C., Seitz, D., Sheppard, D.A.,
Siy, A., Stevens, K., Summers, D., Szabo, A., Timofeeva, M.,
Vaivads, A., Velli, M., Yehle, A., Werthimer, D., Wygant, J.R.: 2016,
The FIELDS instrument suite for Solar Probe Plus: Measuring the
coronal plasma and magnetic field, plasma waves and turbulence, and
radio signatures of solar transients.
\textit{Space Sci.\  Rev.} \textbf{204}, 49.
\href{https://doi.org/10.1007/s11214-016-0244-5}{DOI}.
\href{https://ui.adsabs.harvard.edu/abs/2016SSRv..204...49B/abstract}{ADS}.

\bibitem[\protect\citeauthoryear{{Bale et al.}}{2021}]{Ba21}
Bale, S.D., Horbury, T.S., Velli, M., Desai, M.I., Halekas, J.S.,
McManus, M.D., Panasenco, O., Badman, S.T., Bowen, T.A., Chandran, B.D.G.,
Drake, J.F., Kasper, J.C., Laker, R., Mallet, A., Matteini, L., Phan, T.D.,
Raouafi, N.E., Squire, J., Woodham, L.D., Woolley, T.: 2021,
A solar source of Alfv\'{e}nic magnetic field switchbacks: In situ
remnants of magnetic funnels on supergranulation scales.
\textit{Astrophys.\  J.} \textbf{923}, 174.
\href{https://doi.org/10.3847/1538-4357/ac2d8c}{DOI}.
\href{https://ui.adsabs.harvard.edu/abs/2021ApJ...923..174B/abstract}{ADS}.

\bibitem[\protect\citeauthoryear{{Bandyopadhyay et al.}}{2022}]{By22}
Bandyopadhyay, R., Matthaeus, W.H., McComas, D.J., Chhiber, R.,
Usmanov, A.V., Huang, J., Livi, R., Larson, D.E., Kasper, J.C.,
Case, A.W., Stevens, M., Whittlesey, P., Romeo, O.M., Bale, S.D.,
Bonnell, J.W., Dudok de Wit, T., Goetz, K., Harvey, P.R.,
MacDowall, R.J., Malaspina, D.M., Pulupa, M.: 2022,
Sub-Alfv\'{e}nic solar wind observed by the Parker Solar Probe:
Characterization of turbulence, anisotropy, intermittency, and switchback.
\textit{Astrophys.\  J.\  Lett.} \textbf{926}, L1.
\href{https://doi.org/10.3847/2041-8213/ac4a5c}{DOI}.
\href{https://ui.adsabs.harvard.edu/abs/2022ApJ...926L...1B/abstract}{ADS}.

\bibitem[\protect\citeauthoryear{{Barkhudarov}}{1991}]{Bk91}
Barkhudarov, M.R.: 1991,
Alfv\'{e}n waves in stellar winds.
\textit{Solar Phys.} \textbf{135}, 131.
\href{https://doi.org/10.1007/BF00146703}{DOI}.
\href{https://ui.adsabs.harvard.edu/abs/1991SoPh..135..131B/abstract}{ADS}.

\bibitem[\protect\citeauthoryear{{Barnes}}{2007}]{Ba07}
Barnes, S.A.: 2007,
Ages for illustrative field stars using gyrochronology: Viability,
limitations, and errors.
\textit{Astrophys.\  J.} \textbf{669}, 1167.
\href{https://doi.org/10.1086/519295}{DOI}.
\href{https://ui.adsabs.harvard.edu/abs/2007ApJ...669.1167B/abstract}{ADS}.

\bibitem[\protect\citeauthoryear{{Belcher}}{1971}]{B71}
Belcher, J.W.: 1971,
Alfv\'{e}nic wave pressures and the solar wind.
\textit{Astrophys.\  J.} \textbf{168}, 509.
\href{https://doi.org/10.1086/151105}{DOI}.
\href{https://ui.adsabs.harvard.edu/abs/1971ApJ...168..509B/abstract}{ADS}.

\bibitem[\protect\citeauthoryear{{Belcher and MacGregor}}{1976}]{BM76}
Belcher, J.W., MacGregor, K.B.: 1976,
Magnetic acceleration of winds from solar-type stars.
\textit{Astrophys.\  J.} \textbf{210}, 498.
\href{https://doi.org/10.1086/154853}{DOI}.
\href{https://ui.adsabs.harvard.edu/abs/1976ApJ...210..498B/abstract}{ADS}.

\bibitem[\protect\citeauthoryear{{Borovsky}}{2008}]{Bv08}
Borovsky, J.E.: 2008,
Flux tube texture of the solar wind: Strands of the magnetic carpet
at 1 AU?
\textit{J.\  Geophys.\  Res.} \textbf{113}, A08110.
\href{https://doi.org/10.1029/2007JA012684}{DOI}.
\href{https://ui.adsabs.harvard.edu/abs/2008JGRA..113.8110B/abstract}{ADS}.

\bibitem[\protect\citeauthoryear{{Bourouaine and Perez}}{2018}]{BP18}
Bourouaine, S., Perez, J.C.: 2018,
On the limitations of Taylor's hypothesis in Parker Solar Probe's
measurements near the Alfv\'{e}n critical point.
\textit{Astrophys.\  J.\  Lett.} \textbf{858}, L20.
\href{https://doi.org/10.3847/2041-8213/aabccf}{DOI}.
\href{https://ui.adsabs.harvard.edu/abs/2018ApJ...858L..20B/abstract}{ADS}.

\bibitem[\protect\citeauthoryear{{Bretherton and Garrett}}{1968}]{BG68}
Bretherton, F.P., Garrett, C.J.R.: 1968,
Wavetrains in inhomogeneous moving media.
\textit{Proc.\  Roy.\  Soc.\  London A} \textbf{302}, 529.
\href{https://doi.org/10.1098/rspa.1968.0034}{DOI}.
\href{https://ui.adsabs.harvard.edu/abs/1968RSPSA.302..529B/abstract}{ADS}.

\bibitem[\protect\citeauthoryear{{Case et al.}}{2020}]{Case20}
Case, A.W., Kasper, J.C., Stevens, M.L., Korreck, K.E., Paulson, K.,
Daigneau, P., Caldwell, D., Freeman, M., Henry, T., Klingensmith, B.,
Bookbinder, J.A., Robinson, M., Berg, P., Tiu, C., Wright, K.H.,
Reinhart, M.J., Curtis, D., Ludlam, M., Larson, D., Whittlesey, P.,
Livi, R., Klein, K.G., Martinovi\'{c}, M.M.: 2020,
The Solar Probe Cup on the Parker Solar Probe.
\textit{Astrophys.\  J.\  Suppl.} \textbf{246}, 43.
\href{https://doi.org/10.3847/1538-4365/ab5a7b}{DOI}.
\href{https://ui.adsabs.harvard.edu/abs/2020ApJS..246...43C/abstract}{ADS}.

\bibitem[\protect\citeauthoryear{{Chae and Sakurai}}{2008}]{CS08}
Chae, J., Sakurai, T.: 2008,
A test of three optical flow techniques: LCT, DAVE, and NAVE.
\textit{Astrophys.\  J.} \textbf{689}, 593.
\href{https://doi.org/10.1086/592761}{DOI}.
\href{https://ui.adsabs.harvard.edu/abs/2008ApJ...689..593C/abstract}{ADS}.

\bibitem[\protect\citeauthoryear{{Chandran and Hollweg}}{2009}]{CH09}
Chandran, B.D.G., Hollweg, J.V.: 2009,
Alfv\'{e}n wave reflection and turbulent heating in the solar
wind from 1 solar radius to 1 AU: An analytical treatment.
\textit{Astrophys.\  J.} \textbf{707}, 1659.
\href{https://doi.org/10.1088/0004-637X/707/2/1659}{DOI}.
\href{https://ui.adsabs.harvard.edu/abs/2009ApJ...707.1659C/abstract}{ADS}.

\bibitem[\protect\citeauthoryear{{Chandran and Perez}}{2019}]{CP19}
Chandran, B.D.G., Perez, J.C.: 2019,
Reflection-driven magnetohydrodynamic turbulence in the solar
atmosphere and solar wind.
\textit{J.\  Plasma Phys.} \textbf{85}, 905850409.
\href{https://doi.org/10.1017/S0022377819000540}{DOI}.
\href{https://ui.adsabs.harvard.edu/abs/2019JPlPh..85d9009C/abstract}{ADS}.

\bibitem[\protect\citeauthoryear{{Chhiber}}{2022}]{Ch22solo}
Chhiber, R.: 2022,
Anisotropic magnetic turbulence in the inner heliosphere: Radial
evolution of distributions observed by Parker Solar Probe.
\textit{Astrophys.\  J.} \textbf{939}, 33.
\href{https://doi.org/10.3847/1538-4357/ac9386}{DOI}.
\href{https://ui.adsabs.harvard.edu/abs/2022ApJ...939...33C/abstract}{ADS}.

\bibitem[\protect\citeauthoryear{{Chhiber et al.}}{2022}]{Ch22}
Chhiber, R., Matthaeus, W.H., Usmanov, A.V., Bandyopadhyay, R.,
Goldstein, M.L.: 2022,
An extended and fragmented Alfv\'{e}n zone in the young solar wind.
\textit{Mon.\  Not.\  Roy.\  Astron.\  Soc.} \textbf{513}, 159.
\href{https://doi.org/10.1093/mnras/stac779}{DOI}.
\href{https://ui.adsabs.harvard.edu/abs/2022MNRAS.513..159C/abstract}{ADS}.

\bibitem[\protect\citeauthoryear{{Chhiber et al.}}{2019}]{Ch19}
Chhiber, R., Usmanov, A.V., Matthaeus, W.H., Goldstein, M.L.: 2019,
Contextual predictions for the Parker Solar Probe, I:
Critical surfaces and regions.
\textit{Astrophys.\  J.\  Suppl.} \textbf{241}, 11.
\href{https://doi.org/10.3847/1538-4365/ab0652}{DOI}.
\href{https://ui.adsabs.harvard.edu/abs/2019ApJS..241...11C/abstract}{ADS}.

\bibitem[\protect\citeauthoryear{{Cohen}}{2015}]{Co15}
Cohen, O.: 2015,
Quantifying the difference between the flux-tube expansion factor at
the source surface and at the Alfv\'{e}n surface using a global MHD
model for the solar wind.
\textit{Solar Phys.} \textbf{290}, 2245.
\href{https://doi.org/10.1007/s11207-015-0739-3}{DOI}.
\href{https://ui.adsabs.harvard.edu/abs/2015SoPh..290.2245C/abstract}{ADS}.

\bibitem[\protect\citeauthoryear{{Cohen et al.}}{2009}]{Co09}
Cohen, O., Drake, J.J., Kashyap, V.L., Gombosi, T.I.: 2009,
The effect of magnetic spots on stellar winds and angular momentum loss.
\textit{Astrophys.\  J.} \textbf{699}, 1501.
\href{https://doi.org/10.1088/0004-637X/699/2/1501}{DOI}.
\href{https://ui.adsabs.harvard.edu/abs/2009ApJ...699.1501C/abstract}{ADS}.

\bibitem[\protect\citeauthoryear{{Colaninno and Vourlidas}}{2006}]{CV06}
Colaninno, R.C., Vourlidas, A.: 2006,
Analysis of the velocity field of CMEs using optical flow methods.
\textit{Astrophys.\  J.} \textbf{652}, 1747.
\href{https://doi.org/10.1086/507943}{DOI}.
\href{https://ui.adsabs.harvard.edu/abs/2006ApJ...652.1747C/abstract}{ADS}.

\bibitem[\protect\citeauthoryear{{Cranmer}}{2010}]{Cr10}
Cranmer, S.R.: 2010,
An efficient approximation of the coronal heating rate for use in
global Sun-heliosphere simulations.
\textit{Astrophys.\  J.} \textbf{710}, 676.
\href{https://doi.org/10.1088/0004-637X/710/1/676}{DOI}.
\href{https://ui.adsabs.harvard.edu/abs/2010ApJ...710..676C/abstract}{ADS}.

\bibitem[\protect\citeauthoryear{{Cranmer}}{2020}]{Cr20}
Cranmer, S.R.: 2020,
Updated measurements of proton, electron, and oxygen temperatures
in the fast solar wind.
\textit{Res.\  Notes Am.\  Astron.\  Soc.} \textbf{4}, 249.
\href{https://doi.org/10.3847/2515-5172/abd5ae}{DOI}.
\href{https://ui.adsabs.harvard.edu/abs/2020RNAAS...4..249C/abstract}{ADS}.

\bibitem[\protect\citeauthoryear{{Cranmer et al.}}{2021}]{Cr21}
Cranmer, S.R., DeForest, C.E., Gibson, S.E.: 2021,
Inward-propagating plasma parcels in the solar corona: Models with
aerodynamic drag, ablation, and snowplow accretion.
\textit{Astrophys.\  J.} \textbf{913}, 4.
\href{https://doi.org/10.3847/1538-4357/abf146}{DOI}.
\href{https://ui.adsabs.harvard.edu/abs/2021ApJ...913....4C/abstract}{ADS}.

\bibitem[\protect\citeauthoryear{{Cranmer et al.}}{2007}]{CvB07}
Cranmer, S.R., van Ballegooijen, A.A., Edgar, R.J.: 2007,
Self-consistent coronal heating and solar wind acceleration from
anisotropic magnetohydrodynamic turbulence.
\textit{Astrophys.\  J.\  Suppl.} \textbf{171}, 520.
\href{https://doi.org/10.1086/518001}{DOI}.
\href{https://ui.adsabs.harvard.edu/abs/2007ApJS..171..520C/abstract}{ADS}.

\bibitem[\protect\citeauthoryear{{Cranmer et al.}}{2013}]{CvB13}
Cranmer, S.R., van Ballegooijen, A.A., Woolsey, L.N.: 2013,
Connecting the Sun's high-resolution magnetic carpet to the
turbulent heliosphere.
\textit{Astrophys.\  J.} \textbf{767}, 125.
\href{https://doi.org/10.1088/0004-637X/767/2/125}{DOI}.
\href{https://ui.adsabs.harvard.edu/abs/2013ApJ...767..125C/abstract}{ADS}.

\bibitem[\protect\citeauthoryear{{Crooker et al.}}{2002}]{Ck02}
Crooker, N.U., Gosling, J.T., Kahler, S.W.: 2002,
Reducing heliospheric magnetic flux from coronal mass ejections
without disconnection.
\textit{J.\  Geophys.\  Res.} \textbf{107}, 1028.
\href{https://doi.org/10.1029/2001JA000236}{DOI}.
\href{https://ui.adsabs.harvard.edu/abs/2002JGRA..107.1028C/abstract}{ADS}.

\bibitem[\protect\citeauthoryear{{Dalla and Fletcher}}{2016}]{DF16}
Dalla, S., Fletcher, L.: 2016,
A pioneer of solar astronomy.
\textit{Astron.\  Geophys.} \textbf{57}, 5.21.
\href{https://doi.org/10.1093/astrogeo/atw181}{DOI}.
\href{https://ui.adsabs.harvard.edu/abs/2016A&G....57e5.21D/abstract}{ADS}.

\bibitem[\protect\citeauthoryear{{DeForest et al.}}{2017}]{De17}
DeForest, C.E., de Koning, C.A., Elliott, H.A.: 2017,
3D polarized imaging of coronal mass ejections: Chirality of a CME.
\textit{Astrophys.\  J.} \textbf{850}, 130.
\href{https://doi.org/10.3847/1538-4357/aa94ca}{DOI}.
\href{https://ui.adsabs.harvard.edu/abs/2017ApJ...850..130D/abstract}{ADS}.

\bibitem[\protect\citeauthoryear{{DeForest et al.}}{2012}]{De12}
DeForest, C.E., Howard, T.A., McComas, D.J.: 2012,
Disconnecting open solar magnetic flux.
\textit{Astrophys.\  J.} \textbf{745}, 36.
\href{https://doi.org/10.1088/0004-637X/745/1/36}{DOI}.
\href{https://ui.adsabs.harvard.edu/abs/2012ApJ...745...36D/abstract}{ADS}.

\bibitem[\protect\citeauthoryear{{DeForest et al.}}{2014}]{De14}
DeForest, C.E., Howard, T.A., McComas, D.J.: 2014,
Inbound waves in the solar corona: A direct indicator of the
Alfv\'{e}n surface location.
\textit{Astrophys.\  J.} \textbf{787}, 124.
\href{https://doi.org/10.1088/0004-637X/787/2/124}{DOI}.
\href{https://ui.adsabs.harvard.edu/abs/2014ApJ...787..124D/abstract}{ADS}.

\bibitem[\protect\citeauthoryear{{DeForest et al.}}{2018}]{De18}
DeForest, C.E., Howard, R.A., Velli, M., Viall, N., Vourlidas, A.: 2018,
The highly structured outer solar corona.
\textit{Astrophys.\  J.} \textbf{862}, 18.
\href{https://doi.org/10.3847/1538-4357/aac8e3}{DOI}.
\href{https://ui.adsabs.harvard.edu/abs/2018ApJ...862...18D/abstract}{ADS}.

\bibitem[\protect\citeauthoryear{{DeForest et al.}}{2022}]{De22}
DeForest, C.E., Killough, R., Gibson, S., Henry, A., Case, T.,
Beasley, M., Laurent, G., Colaninno, R.,
Waltham, N., PUNCH Science Team: 2022,
Polarimeter to Unify the Corona and Heliosphere (PUNCH): Science,
status, and path to flight.
In: \textit{Proc.\  2022 IEEE Aerospace Conference}, 1.
\href{https://doi.org/10.1109/AERO53065.2022.9843340}{DOI}.
\href{https://ui.adsabs.harvard.edu/abs/2022aero.confE...1D/abstract}{ADS}.

\bibitem[\protect\citeauthoryear{{DeForest et al.}}{2016}]{De16}
DeForest, C.E., Matthaeus, W.H., Viall, N.M., Cranmer, S.R.: 2016,
Fading coronal structure and the onset of turbulence in the
young solar wind.
\textit{Astrophys.\  J.} \textbf{828}, 66.
\href{https://doi.org/10.3847/0004-637X/828/2/66}{DOI}.
\href{https://ui.adsabs.harvard.edu/abs/2016ApJ...828...66D/abstract}{ADS}.

\bibitem[\protect\citeauthoryear{{Dmitruk et al.}}{2002}]{Dm02}
Dmitruk, P., Matthaeus, W.H., Milano, L.J., Oughton, S., Zank, G.P.,
Mullan, D.J.: 2002,
Coronal heating distribution due to low-frequency, wave-driven turbulence.
\textit{Astrophys.\  J.} \textbf{575}, 571.
\href{https://doi.org/10.1086/341188}{DOI}.
\href{https://ui.adsabs.harvard.edu/abs/2002ApJ...575..571D/abstract}{ADS}.

\bibitem[\protect\citeauthoryear{{Durney and Stenflo}}{1972}]{DS72}
Durney, B.R., Stenflo, J.O.: 1972,
On stellar activity cycles.
\textit{Astrophys.\  Space Sci.} \textbf{15}, 307.
\href{https://doi.org/10.1007/BF00649924}{DOI}.
\href{https://ui.adsabs.harvard.edu/abs/1972Ap&SS..15..307D/abstract}{ADS}.

\bibitem[\protect\citeauthoryear{{Elsasser}}{1950}]{E50}
Elsasser, W.M.: 1950, The hydromagnetic equations.
\textit{Phys.\  Rev.} \textbf{79}, 183.
\href{https://doi.org/10.1103/PhysRev.79.183}{DOI}.
\href{https://ui.adsabs.harvard.edu/abs/1950PhRv...79..183E/abstract}{ADS}.

\bibitem[\protect\citeauthoryear{{Exarhos and Moussas}}{2000}]{EM00}
Exarhos, G., Moussas, X.: 2000,
An estimation of the shape and temporal variation of the solar wind
sonic, Alfv\'{e}nic, and fast magnetosonic surfaces.
\textit{Astron.\  Astrophys.} \textbf{356}, 315.
\href{https://ui.adsabs.harvard.edu/abs/2000A&A...356..315E/abstract}{ADS}.

\bibitem[\protect\citeauthoryear{{Ferraro}}{1937}]{F37}
Ferraro, V.C.A.: 1937,
The non-uniform rotation of the Sun and its magnetic field.
\textit{Mon.\  Not.\  Roy.\  Astron.\  Soc.} \textbf{97}, 458.
\href{https://doi.org/10.1093/mnras/97.6.458}{DOI}.
\href{https://ui.adsabs.harvard.edu/abs/1937MNRAS..97..458F/abstract}{ADS}.

\bibitem[\protect\citeauthoryear{{Finley and Brun}}{2023}]{Fi23}
Finley, A.J., Brun, A.S.: 2023,
Accounting for differential rotation in calculations of the Sun's
angular momentum-loss rate.
\textit{Astron.\  Astrophys.} \textbf{674}, A42.
\href{https://doi.org/10.1051/0004-6361/202245642}{DOI}.
\href{https://ui.adsabs.harvard.edu/abs/2023A&A...674A..42F/abstract}{ADS}.

\bibitem[\protect\citeauthoryear{{Finley et al.}}{2019}]{Fi19}
Finley, A.J., Deshmukh, S., Matt, S.P., Owens, M., Wu, C.-J.: 2019,
Solar angular momentum loss over the past several millennia.
\textit{Astrophys.\  J.} \textbf{883}, 67.
\href{https://doi.org/10.3847/1538-4357/ab3729}{DOI}.
\href{https://ui.adsabs.harvard.edu/abs/2019ApJ...883...67F/abstract}{ADS}.

\bibitem[\protect\citeauthoryear{{Fox et al.}}{2016}]{Fo16}
Fox, N.J., Velli, M.C., Bale, S.D., Decker, R., Driesman, A.,
Howard, R.A., Kasper, J.C., Kinnison, J., Kusterer, M., Lario, D.,
Lockwood, M.K., McComas, D.J., Raouafi, N.E., Szabo, A.: 2016,
The Solar Probe Plus mission: Humanity's first visit to our star.
\textit{Space Sci.\  Rev.} \textbf{204}, 7.
\href{https://doi.org/10.1007/s11214-015-0211-6}{DOI}.
\href{https://ui.adsabs.harvard.edu/abs/2016SSRv..204....7F/abstract}{ADS}.

\bibitem[\protect\citeauthoryear{{Galano et al.}}{2018}]{Ga18}
Galano, D., Bemporad, A., Buckley, S., Cernica, I., D\'{a}niel, V.,
Denis, F., de Vos, L., Fineschi, S., Galy, C., Graczyk, R., Horodyska, P.,
Jacob, J., Jansen, R., Kranitis, N., Kurowski, M., Ladno, M.,
Ledent, P., Loreggia, D., Melich, R., Mollet, D., Mosdorf, M.,
Paschalis, A., Peresty, R., Purica, M., Radzik, B., Rataj, M.,
Rougeot, R., Salvador, L., Thizy, C., Versluys, J., Walczak, T.,
Zarzycka, A., Zender, J., Zhukov, A.: 2018,
Development of ASPIICS: A coronagraph based on Proba-3 formation
flying mission.
\textit{Proc.\  SPIE} \textbf{10698}, 106982Y.
\href{https://doi.org/10.1117/12.2312493}{DOI}.
\href{https://ui.adsabs.harvard.edu/abs/2018SPIE10698E..2YG/abstract}{ADS}.

\bibitem[\protect\citeauthoryear{{Garcia-Sage et al.}}{2023}]{GS23}
Garcia-Sage, K., Farrish, A.O., Airapetian, V.S., Alexander, D., Cohen, O.,
Domagal-Goldman, S., Dong, C., Gronoff, G., Halford, A.J., Lazio, J.,
Luhmann, J.G., Schwieterman, E., Sciola, A., Segura, A., Toffoletto, F.,
Vievering, J., Ahmed, M.R., Bali, K., Rau, G.: 2023,
Star-exoplanet interactions: A growing interdisciplinary field in
heliophysics.
\textit{Front.\  Astron.\  Space Sci.} \textbf{10}, 1064076.
\href{https://doi.org/10.3389/fspas.2023.1064076}{DOI}.
\href{https://ui.adsabs.harvard.edu/abs/2023FrASS..1064076G/abstract}{ADS}.

\bibitem[\protect\citeauthoryear{{Garraffo et al.}}{2017}]{Gf17}
Garraffo, C., Drake, J.J., Cohen, O., Alvarado-G\'{o}mez, J.D.,
Moschou, S.P.: 2017,
The threatening magnetic and plasma environment of the TRAPPIST-1 planets.
\textit{Astrophys.\  J.\  Lett.} \textbf{843}, L33.
\href{https://doi.org/10.3847/2041-8213/aa79ed}{DOI}.
\href{https://ui.adsabs.harvard.edu/abs/2017ApJ...843L..33G/abstract}{ADS}.

\bibitem[\protect\citeauthoryear{{Gibson et al.}}{2016}]{Gi16}
Gibson, S., Kucera, T., White, S., Dove, J., Fan, Y., Forland, B.,
Rachmeler, L., Downs, C., Reeves, K.: 2016,
FORWARD: A toolset for multiwavelength coronal magnetometry.
\textit{Front.\  Astron.\  Space Sci.} \textbf{3}, 8.
\href{https://doi.org/10.3389/fspas.2016.00008}{DOI}.
\href{https://ui.adsabs.harvard.edu/abs/2016FrASS...3....8G/abstract}{ADS}.

\bibitem[\protect\citeauthoryear{{Gilly et al.}}{2021}]{GC21}
Gilly, C.R., Cranmer, S.R., Gibson, S.E.: 2021,
STRIA: A new module within FORWARD towards modelling PUNCH datasets.
\textit{Bull.\  Am.\  Astron.\  Soc.}, \textbf{53}, 2021n6i328p02.
\href{https://ui.adsabs.harvard.edu/abs/2021AAS...23832802G/abstract}{ADS}.

\bibitem[\protect\citeauthoryear{{Goelzer et al.}}{2014}]{Go14}
Goelzer, M.L., Schwadron, N.A., Smith, C.W.: 2014,
An analysis of Alfv\'{e}n radius based on sunspot number from 1749 to today.
\textit{J.\  Geophys.\  Res.} \textbf{119}, 115.
\href{https://doi.org/10.1002/2013JA019420}{DOI}.
\href{https://ui.adsabs.harvard.edu/abs/2014JGRA..119..115G/abstract}{ADS}.

\bibitem[\protect\citeauthoryear{{Gombosi et al.}}{2018}]{Gm18}
Gombosi, T.I., van der Holst, B., Manchester, W.B., Sokolov, I.V.: 2018,
Extended MHD modeling of the steady solar corona and the solar wind.
\textit{Liv.\  Rev.\  Solar Phys.} \textbf{15}, 4.
\href{https://doi.org/10.1007/s41116-018-0014-4}{DOI}.
\href{https://ui.adsabs.harvard.edu/abs/2018LRSP...15....4G/abstract}{ADS}.

\bibitem[\protect\citeauthoryear{{Gosling et al.}}{1987}]{Go87}
Gosling, J.T., Baker, D.N., Bame, S.J., Feldman, W.C., Zwickl, R.D.,
Smith, E.J.: 1987,
Bidirectional solar wind electron heat flux events.
\textit{J.\  Geophys.\  Res.} \textbf{92}, 8519.
\href{https://doi.org/10.1029/JA092iA08p08519}{DOI}.
\href{https://ui.adsabs.harvard.edu/abs/1987JGR....92.8519G/abstract}{ADS}.

\bibitem[\protect\citeauthoryear{{Griton et al.}}{2020}]{Gr20}
Griton, L., Pinto, R.F., Poirier, N., Kouloumvakos, A., Lavarra, M.,
Rouillard, A.P.: 2020,
Coronal bright points as possible sources of density variations in
the solar corona.
\textit{Astrophys.\  J.} \textbf{893}, 64.
\href{https://doi.org/10.3847/1538-4357/ab7b76}{DOI}.
\href{https://ui.adsabs.harvard.edu/abs/2020ApJ...893...64G/abstract}{ADS}.

\bibitem[\protect\citeauthoryear{{Harding}}{2013}]{Ha13}
Harding, A. K.: 2013,
The neutron star zoo.
\textit{Front.\  Phys.} \textbf{8}, 679.
\href{https://doi.org/10.1007/s11467-013-0285-0}{DOI}.
\href{https://ui.adsabs.harvard.edu/abs/2013FrPhy...8..679H/abstract}{ADS}.

\bibitem[\protect\citeauthoryear{{Heinemann and Olbert}}{1980}]{HO80}
Heinemann, M., Olbert, S.: 1980,
Non-WKB Alfv\'{e}n waves in the solar wind.
\textit{J.\  Geophys.\  Res.} \textbf{85}, 1311.
\href{https://doi.org/10.1029/JA085iA03p01311}{DOI}.
\href{https://ui.adsabs.harvard.edu/abs/1980JGR....85.1311H/abstract}{ADS}.

\bibitem[\protect\citeauthoryear{{Hollweg}}{1970}]{Ho70}
Hollweg, J.V.: 1970,
Collisionless solar wind, I: Constant electron temperature.
\textit{J.\  Geophys.\  Res.} \textbf{75}, 2403.
\href{https://doi.org/10.1029/JA075i013p02403}{DOI}.
\href{https://ui.adsabs.harvard.edu/abs/1970JGR....75.2403H/abstract}{ADS}.

\bibitem[\protect\citeauthoryear{{Holzer and Leer}}{1997}]{HL97}
Holzer, T.E., Leer, E.: 1997,
Coronal hole structure and the high-speed solar wind.
In: A. Wilson (ed.), \textit{Fifth SOHO Workshop: The Corona and
Solar Wind Near Minimum Activity,} \textbf{404}, ESA, Noordwijk, 65.
\href{https://ui.adsabs.harvard.edu/abs/1997ESASP.404...65H/abstract}{ADS}.

\bibitem[\protect\citeauthoryear{{Hossain et al.}}{1995}]{Hs95}
Hossain, M., Gray, P.C., Pontius, D.H., Matthaeus, W.H., Oughton, S.: 1995,
Phenomenology for the decay of energy-containing eddies in
homogeneous MHD turbulence.
\textit{Phys.\  Fluids} \textbf{7}, 2886.
\href{https://doi.org/10.1063/1.868665}{DOI}.
\href{https://ui.adsabs.harvard.edu/abs/1995PhFl....7.2886H/abstract}{ADS}.

\bibitem[\protect\citeauthoryear{{Howard et al.}}{2008}]{How08}
Howard, R.A., Moses, J.D., Vourlidas, A., Newmark, J.S., Socker, D.G.,
Plunkett, S.P., Korendyke, C.M., Cook, J.W., Hurley, A., Davila, J.M.,
Thompson, W.T., St Cyr, O.C., Mentzell, E., Mehalick, K., Lemen, J.R.,
Wuelser, J.P., Duncan, D.W., Tarbell, T.D., Wolfson, C.J., Moore, A.,
Harrison, R.A., Waltham, N.R., Lang, J., Davis, C.J., Eyles, C.J.,
Mapson-Menard, H., Simnett, G.M., Halain, J.P., Defise, J.M., Mazy, E.,
Rochus, P., Mercier, R., Ravet, M.F., Delmotte, F., Auchere, F.,
Delaboudiniere, J.P., Bothmer, V., Deutsch, W., Wang, D., Rich, N.,
Cooper, S., Stephens, V., Maahs, G., Baugh, R., McMullin, D.,
Carter, T.: 2008,
Sun Earth Connection Coronal and Heliospheric Investigation (SECCHI).
\textit{Space Sci.\  Rev.} \textbf{136}, 67.
\href{https://doi.org/10.1007/s11214-008-9341-4}{DOI}.
\href{https://ui.adsabs.harvard.edu/abs/2008SSRv..136...67H/abstract}{ADS}.

\bibitem[\protect\citeauthoryear{{Hundhausen}}{1972}]{Hu72}
Hundhausen, A.J.: 1972,
\textit{Coronal Expansion and Solar Wind,
Phys.\  Chem.\  in Space,} \textbf{5}, Springer-Verlag, Berlin.
\href{https://doi.org/10.1007/978-3-642-65414-5}{DOI}.
\href{https://ui.adsabs.harvard.edu/abs/1972cesw.book.....H/abstract}{ADS}.

\bibitem[\protect\citeauthoryear{{Iroshnikov}}{1963}]{Ir63}
Iroshnikov, P.S.: 1963,
Turbulence of a conducting fluid in a strong magnetic field.
\textit{Astron.\  Zh.} \textbf{40}, 742.
\href{https://ui.adsabs.harvard.edu/abs/1963AZh....40..742I/abstract}{ADS}.

\bibitem[\protect\citeauthoryear{{Isenberg and Hollweg}}{1982}]{IH82}
Isenberg, P.A., Hollweg, J.V.: 1982,
Finite amplitude Alfv\'{e}n waves in a multi-ion plasma: Propagation,
acceleration, and heating.
\textit{J.\  Geophys.\  Res.} \textbf{87}, 5023.
\href{https://doi.org/10.1029/JA087iA07p05023}{DOI}.
\href{https://ui.adsabs.harvard.edu/abs/1982JGR....87.5023I/abstract}{ADS}.

\bibitem[\protect\citeauthoryear{{Jacques}}{1977}]{J77}
Jacques, S.A.: 1977,
Momentum and energy transport by waves in the solar atmosphere
and solar wind.
\textit{Astrophys.\  J.} \textbf{215}, 942.
\href{https://doi.org/10.1086/155430}{DOI}.
\href{https://ui.adsabs.harvard.edu/abs/1977ApJ...215..942J/abstract}{ADS}.

\bibitem[\protect\citeauthoryear{{Kasper et al.}}{2016}]{Ka16}
Kasper, J.C., Abiad, R., Austin, G., Balat-Pichelin, M.,
Bale, S.D., Belcher, J.W., Berg, P., Bergner, H.,
Berthomier, M., Bookbinder, J., Brodu, E., Caldwell, D.,
Case, A.W., Chandran, B.D.G., Cheimets, P.,
Cirtain, J.W., Cranmer, S.R., Curtis, D.W., Daigneau, P.,
Dalton, G., Dasgupta, B., DeTomaso, D., Diaz-Aguado, M.,
Djordjevic, B., Donaskowski, B., Effinger, M., Florinski, V.,
Fox, N., Freeman, M., Gallagher, D., Gary, S.P., Gauron, T.,
Gates, R., Goldstein, M., Golub, L., Gordon, D.A.,
Gurnee, R., Guth, G., Halekas, J., Hatch, K., Heerikuisen, J.,
Ho, G., Hu, Q., Johnson, G., Jordan, S.P., Korreck, K.E.,
Larson, D., Lazarus, A.J., Li, G., Livi, R., Ludlam, M.,
Maksimovic, M., McFadden, J.P., Marchant, W.,
Maruca, B.A., McComas, D.J., Messina, L., Mercer, T.,
Park, S., Peddie, A.M., Pogorelov, N., Reinhart, M.J.,
Richardson, J.D., Robinson, M., Rosen, I., Skoug, R.M.,
Slagle, A., Steinberg, J.T., Stevens, M.L., Szabo, A.,
Taylor, E.R., Tiu, C., Turin, P., Velli, M., Webb, G.,
Whittlesey, P., Wright, K., Wu, S.T., Zank, G.: 2016,
Solar Wind Electrons Alphas and Protons (SWEAP) investigation:
Design of the solar wind and coronal plasma instrument suite for
Solar Probe Plus.
\textit{Space Sci.\  Rev.} \textbf{204}, 131.
\href{https://doi.org/10.1007/s11214-015-0206-3}{DOI}.
\href{https://ui.adsabs.harvard.edu/abs/2016SSRv..204..131K/abstract}{ADS}.

\bibitem[\protect\citeauthoryear{{Kasper and Klein}}{2019}]{KK19}
Kasper, J.C., Klein, K.G.: 2019,
Strong preferential ion heating is limited to within the solar
Alfv\'{e}n surface.
\textit{Astrophys.\  J.\  Lett.} \textbf{877}, L35.
\href{https://doi.org/10.3847/2041-8213/ab1de5}{DOI}.
\href{https://ui.adsabs.harvard.edu/abs/2019ApJ...877L..35K/abstract}{ADS}.

\bibitem[\protect\citeauthoryear{{Kasper et al.}}{2021}]{Ka21}
Kasper, J.C., Klein, K.G., Lichko, E., Huang, J., Chen, C.H.K.,
Badman, S.T., Bonnell, J., Whittlesey, P.L., Livi, R., Larson, D.,
Pulupa, M., Rahmati, A., Stansby, D., Korreck, K.E., Stevens, M.,
Case, A.W., Bale, S.D., Maksimovic, M., Moncuquet, M., Goetz, K.,
Halekas, J.S., Malaspina, D., Raouafi, N.E., Szabo, A., MacDowall, R.,
Velli, M., Dudok de Wit, T., Zank, G.P.: 2021,
Parker Solar Probe enters the magnetically dominated solar corona.
\textit{Phys.\  Rev.\  Lett.} \textbf{127}, 255101.
\href{https://doi.org/10.1103/PhysRevLett.127.255101}{DOI}.
\href{https://ui.adsabs.harvard.edu/abs/2021PhRvL.127y5101K/abstract}{ADS}.

\bibitem[\protect\citeauthoryear{{Katsikas et al.}}{2010}]{Kt10}
Katsikas, V., Exarhos, G., Moussas, X.: 2010,
Study of the solar slow sonic, Alfv\'{e}n, and fast magnetosonic
transition surfaces.
\textit{Adv.\  Space Res.} \textbf{46}, 382.
\href{https://doi.org/10.1016/j.asr.2010.05.003}{DOI}.
\href{https://ui.adsabs.harvard.edu/abs/2010AdSpR..46..382K/abstract}{ADS}.

\bibitem[\protect\citeauthoryear{{Keppens and Goedbloed}}{2000}]{KG00}
Keppens, R., Goedbloed, J.P.: 2000,
Stellar winds, dead zones, and coronal mass ejections.
\textit{Astrophys.\  J.} \textbf{530}, 1036.
\href{https://doi.org/10.1086/308395}{DOI}.
\href{https://ui.adsabs.harvard.edu/abs/2000ApJ...530.1036K/abstract}{ADS}.

\bibitem[\protect\citeauthoryear{{Keto}}{2020}]{Ke20}
Keto, E.: 2020,
Stability and solution of the time-dependent Bondi-Parker flow.
\textit{Mon.\  Not.\  Roy.\  Astron.\  Soc.} \textbf{493}, 2834.
\href{https://doi.org/10.1093/mnras/staa529}{DOI}.
\href{https://ui.adsabs.harvard.edu/abs/2020MNRAS.493.2834K/abstract}{ADS}.

\bibitem[\protect\citeauthoryear{{King and Papitashvili}}{2005}]{KP05}
King, J.H., Papitashvili, N.E.: 2005,
Solar wind spatial scales in and comparisons of hourly Wind and ACE
plasma and magnetic field data.
\textit{J.\  Geophys.\  Res.} \textbf{110}, A02104.
\href{https://doi.org/10.1029/2004JA010649}{DOI}.
\href{https://ui.adsabs.harvard.edu/abs/2005JGRA..110.2104K/abstract}{ADS}.

\bibitem[\protect\citeauthoryear{{Klein et al.}}{2015}]{Kn15}
Klein, K.G., Perez, J.C., Verscharen, D., Mallet, A.,
Chandran, B.D.G.: 2015,
A modified version of Taylor's hypothesis for Solar Probe Plus
observations.
\textit{Astrophys.\  J.\  Lett.} \textbf{801}, L18.
\href{https://doi.org/10.3847/2041-8205/801/1/L18}{DOI}.
\href{https://ui.adsabs.harvard.edu/abs/2015ApJ...801L..18K/abstract}{ADS}.

\bibitem[\protect\citeauthoryear{{Kopp and Holzer}}{1976}]{KH76}
Kopp, R.A., Holzer, T.E.: 1976,
Dynamics of coronal hole regions, I: Steady polytropic flows with
multiple critical points.
\textit{Solar Phys.} \textbf{49}, 43.
\href{https://doi.org/10.1007/BF00221484}{DOI}.
\href{https://ui.adsabs.harvard.edu/abs/1976SoPh...49...43K/abstract}{ADS}.

\bibitem[\protect\citeauthoryear{{Kraichnan}}{1965}]{Kr65}
Kraichnan, R.H.: 1965,
Inertial-range spectrum of hydromagnetic turbulence.
\textit{Phys.\  Fluids} \textbf{8}, 1385.
\href{https://doi.org/10.1063/1.1761412}{DOI}.
\href{https://ui.adsabs.harvard.edu/abs/1965PhFl....8.1385K/abstract}{ADS}.

\bibitem[\protect\citeauthoryear{{Li and Li}}{2006}]{LL06}
Li, B., Li, X.: 2006,
Effects of $\alpha$ particles on the angular momentum loss from the Sun.
\textit{Astron.\  Astrophys.} \textbf{456}, 359.
\href{https://doi.org/10.1051/0004-6361:20054624}{DOI}.
\href{https://ui.adsabs.harvard.edu/abs/2006A&A...456..359L/abstract}{ADS}.

\bibitem[\protect\citeauthoryear{{Li}}{1999}]{Lj99}
Li, J.: 1999,
Magnetic braking of the present Sun.
\textit{Mon.\  Not.\  Roy.\  Astron.\  Soc.} \textbf{302}, 203.
\href{https://doi.org/10.1046/j.1365-8711.1999.02134.x}{DOI}.
\href{https://ui.adsabs.harvard.edu/abs/1999MNRAS.302..203L/abstract}{ADS}.

\bibitem[\protect\citeauthoryear{{Li}}{1999}]{Lj98}
Li, J., Raymond, J.C., Acton, L.W., Kohl, J.L., Romoli, M., Noci, G.,
Naletto, G.: 1998,
Physical structure of a coronal streamer in the closed-field region as
observed from UVCS/SOHO and SXT/Yohkoh.
\textit{Astrophys.\  J.} \textbf{506}, 431.
\href{https://doi.org/10.1086/306244}{DOI}.
\href{https://ui.adsabs.harvard.edu/abs/1998ApJ...506..431L/abstract}{ADS}.

\bibitem[\protect\citeauthoryear{{Liewer et al.}}{2004}]{Lw04}
Liewer, P.C., Neugebauer, M., Zurbuchen, T.: 2004,
Characteristics of active-region sources of solar wind near solar maximum.
\textit{Solar Phys.} \textbf{223}, 209.
\href{https://doi.org/10.1007/s11207-004-1105-z}{DOI}.
\href{https://ui.adsabs.harvard.edu/abs/2004SoPh..223..209L/abstract}{ADS}.

\bibitem[\protect\citeauthoryear{{Lifschitz and Goedbloed}}{1997}]{LG97}
Lifschitz, A., Goedbloed, J.P.: 1997,
Transonic magnetohydrodynamic flows.
\textit{J.\  Plasma Phys.} \textbf{58}, 61.
\href{https://doi.org/10.1017/S0022377897005680}{DOI}.
\href{https://ui.adsabs.harvard.edu/abs/1997JPlPh..58...61L/abstract}{ADS}.

\bibitem[\protect\citeauthoryear{{Linker et al.}}{1999}]{Ln99}
Linker, J.A., Miki\'{c}, Z., Biesecker, D.A., Forsyth, R.J.,
Gibson, S.E., Lazarus, A.J., Lecinski, A., Riley, P., Szabo, A.,
Thompson, B.J.: 1999,
Magnetohydrodynamic modeling of the solar corona during Whole Sun Month.
\textit{J.\  Geophys.\  Res.} \textbf{104}, 9809.
\href{https://doi.org/10.1029/1998JA900159}{DOI}.
\href{https://ui.adsabs.harvard.edu/abs/1999JGR...104.9809L/abstract}{ADS}.

\bibitem[\protect\citeauthoryear{{Liu et al.}}{2021}]{Liu21}
Liu, Y.D., Chen, C., Stevens, M.L., Liu, M.: 2021,
Determination of solar wind angular momentum and Alfv\'{e}n radius
from Parker Solar Probe observations.
\textit{Astrophys.\  J.\  Lett.} \textbf{908}, L41.
\href{https://doi.org/10.3847/2041-8213/abe38e}{DOI}.
\href{https://ui.adsabs.harvard.edu/abs/2021ApJ...908L..41L/abstract}{ADS}.

\bibitem[\protect\citeauthoryear{{Liu et al.}}{2023}]{Liu23}
Liu, Y.D., Ran, H., Hu, H., Bale, S.D.: 2023,
On the generation and evolution of switchbacks and the
morphology of the Alfv\'{e}nic transition: Low Mach-number
boundary layers.
\textit{Astrophys.\  J.} \textbf{944}, 116.
\href{https://doi.org/10.3847/1538-4357/acb345}{DOI}.
\href{https://ui.adsabs.harvard.edu/abs/2023ApJ...944..116L/abstract}{ADS}.

\bibitem[\protect\citeauthoryear{{Lloveras et al.}}{2017}]{Ll17}
Lloveras, D.G., V\'{a}squez, A.M., Nuevo, F.A., Frazin, R.A.: 2017,
Comparative study of the three-dimensional thermodynamical structure
of the inner corona of solar minimum Carrington rotations 1915 and 2081.
\textit{Solar Phys.} \textbf{292}, 153.
\href{https://doi.org/10.1007/s11207-017-1179-z}{DOI}.
\href{https://ui.adsabs.harvard.edu/abs/2017SoPh..292..153L/abstract}{ADS}.

\bibitem[\protect\citeauthoryear{{Luhmann et al.}}{2002}]{Lu02}
Luhmann, J.G., Li, Y., Arge, C.N., Gazis, P.R., Ulrich, R.: 2002,
Solar cycle changes in coronal holes and space weather cycles.
\textit{J.\  Geophys.\  Res.} \textbf{107}, 1154.
\href{https://doi.org/10.1029/2001JA007550}{DOI}.
\href{https://ui.adsabs.harvard.edu/abs/2002JGRA..107.1154L/abstract}{ADS}.

\bibitem[\protect\citeauthoryear{{Lynch}}{2020}]{Ly20}
Lynch, B.J.: 2020,
A model for coronal inflows and in/out pairs.
\textit{Astrophys.\  J.} \textbf{905}, 139.
\href{https://doi.org/10.3847/1538-4357/abc5b3}{DOI}.
\href{https://ui.adsabs.harvard.edu/abs/2020ApJ...905..139L/abstract}{ADS}.

\bibitem[\protect\citeauthoryear{{MacGregor and Charbonneau}}{1994}]{MC94}
MacGregor, K.B., Charbonneau, P.: 1994,
Stellar winds with non-WKB Alfv\'{e}n waves, I: Wind models for
solar coronal conditions.
\textit{Astrophys.\  J.} \textbf{430}, 387.
\href{https://doi.org/10.1086/174414}{DOI}.
\href{https://ui.adsabs.harvard.edu/abs/1994ApJ...430..387M/abstract}{ADS}.

\bibitem[\protect\citeauthoryear{{Mackay and Yeates}}{2012}]{MY12}
Mackay, D.H., Yeates, A.R.: 2012,
The Sun's global photospheric and coronal magnetic fields:
Observations and models.
\textit{Liv.\  Rev.\  Solar Phys.} \textbf{9}, 6.
\href{https://doi.org/10.12942/lrsp-2012-6}{DOI}.
\href{https://ui.adsabs.harvard.edu/abs/2012LRSP....9....6M/abstract}{ADS}.

\bibitem[\protect\citeauthoryear{{Marsch and Richter}}{1984}]{MR84}
Marsch, E., Richter, A.K.: 1984,
Distribution of solar wind angular momentum between particles and
magnetic field: Inferences about the Alfv\'{e}n critical point from
Helios observations.
\textit{J.\  Geophys.\  Res.} \textbf{89}, 5386.
\href{https://doi.org/10.1029/JA089iA07p05386}{DOI}.
\href{https://ui.adsabs.harvard.edu/abs/1984JGR....89.5386M/abstract}{ADS}.

\bibitem[\protect\citeauthoryear{{Mart\'{\i}nez Pillet et al.}}{2023}]{MP23}
Mart\'{\i}nez Pillet, V., Cauzzi, G., Tritschler, A., Harra, L., Andretta, A.,
Vourlidas, A., Raouafi, N., Alterman, B.L., Bellot Rubio, L., Cranmer, S.R.,
Gibson, S., De Groof, A., Habbal, S., Ko, Y.-K., Lepri, S.T., Linker, J.,
Malaspina, D.M., Matthews, S., M\"{u}ller, D., Parenti, S., Petrie, G.,
Spadaro, D., Ugarte-Urra, I., Warren, H., Winslow, R., Zouganelis, I.: 2023,
Solar physics in the 2020s: DKIST, Parker Solar Probe, and Solar Orbiter
as a multi-messenger constellation.
In: \textit{The Era of Multi-Messenger Solar Physics,} IAU Symp.\  372,
in press.

\bibitem[\protect\citeauthoryear{{Mason et al.}}{2019}]{Ms19}
Mason, E.I., Antiochos, S.K., Viall, N.M.: 2019,
Observations of solar coronal rain in null point topologies.
\textit{Astrophys.\  J.\  Lett.} \textbf{874}, L33.
\href{https://doi.org/10.3847/2041-8213/ab0c5d}{DOI}.
\href{https://ui.adsabs.harvard.edu/abs/2019ApJ...874L..33M/abstract}{ADS}.

\bibitem[\protect\citeauthoryear{{Matsakos et al.}}{2015}]{Mats15}
Matsakos, T., Uribe, A., K\"{o}nigl, A.: 2015,
Classification of magnetized star-planet interactions: 
Bow shocks, tail, and inspiraling flows.
\textit{Astron.\  Astrophys.} \textbf{578}, A6.
\href{https://doi.org/10.1051/0004-6361/201425593}{DOI}.
\href{https://ui.adsabs.harvard.edu/abs/2015A&A...578A...6M/abstract}{ADS}.

\bibitem[\protect\citeauthoryear{{Matt and Pudritz}}{2005}]{MP05}
Matt, S., Pudritz, R.E.: 2005,
Accretion-powered stellar winds as a solution to the stellar
angular momentum problem.
\textit{Astrophys.\  J.\  Lett.} \textbf{632}, L135.
\href{https://doi.org/10.1086/498066}{DOI}.
\href{https://ui.adsabs.harvard.edu/abs/2005ApJ...632L.135M/abstract}{ADS}.

\bibitem[\protect\citeauthoryear{{Matt and Pudritz}}{2008}]{MP08}
Matt, S., Pudritz, R.E.: 2008,
Accretion-powered stellar winds, II: Numerical solutions for stellar
wind torques.
\textit{Astrophys.\  J.} \textbf{678}, 1109.
\href{https://doi.org/10.1086/533428}{DOI}.
\href{https://ui.adsabs.harvard.edu/abs/2008ApJ...678.1109M/abstract}{ADS}.

\bibitem[\protect\citeauthoryear{{Matthaeus}}{1997}]{Mt97}
Matthaeus, W.H.: 1997,
Turbulence properties along the Solar Probe trajectory.
In: S. Habbal (ed.), \textit{Robotic Exploration Close to the Sun:
Scientific Basis,} \textbf{CP-385}, AIP, Melville, 67.
\href{https://doi.org/10.1063/1.51768}{DOI}.
\href{https://ui.adsabs.harvard.edu/abs/1997AIPC..385...67M/abstract}{ADS}.

\bibitem[\protect\citeauthoryear{{Moraes Filho et al.}}{2022}]{MF22}
Moraes Filho, V., Uritsky, V., Thompson, B. J., DeForest, C.: 2022,
SynCOM: A dynamic model for flow tracking algorithms.
\textit{AGU Fall Meeting}, abs.\  SH12C-1466.
\href{https://ui.adsabs.harvard.edu/abs/2022AGUFMSH12C1466M/abstract}{ADS}.

\bibitem[\protect\citeauthoryear{{Neugebauer and Snyder}}{1962}]{NS62}
Neugebauer, M., Snyder, C.W.: 1962,
Solar plasma experiment.
\textit{Science} \textbf{138}, 1095.
\href{https://doi.org/10.1126/science.138.3545.1095.a}{DOI}.
\href{https://ui.adsabs.harvard.edu/abs/1962Sci...138.1095N/abstract}{ADS}.

\bibitem[\protect\citeauthoryear{{Newmark et al.}}{2020}]{Nw20}
Newmark, J.S., Gopalswamy, N., Kim, Y.H., Viall, N.M., Cho, K.S.F.,
Reginald, N.L., Bong, S.C., Gong, Q., Choi, S., Strachan, L.,
Yashiro, S.: 2020,
The Coronal Diagnostic Experiment (CODEX).
\textit{AGU Fall Meeting}, abstract SH028-0011.
\href{https://ui.adsabs.harvard.edu/abs/2020AGUFMSH0280011N/abstract}{ADS}.

\bibitem[\protect\citeauthoryear{{Owocki}}{2009}]{Ow09}
Owocki, S.P.: 2009,
Stellar magnetospheres.
In: C. Neiner, J.-P. Zahn (eds.),
\textit{Stellar Magnetism, EAS Pub.\  Ser.}, \textbf{39},
EDP, Les Ulis, 223.
\href{https://doi.org/10.1051/eas/0939012}{DOI}.
\href{https://ui.adsabs.harvard.edu/abs/2009EAS....39..223O/abstract}{ADS}.

\bibitem[\protect\citeauthoryear{{Parker}}{1958}]{P58}
Parker, E.N.: 1958,
Dynamics of the interplanetary gas and magnetic fields.
\textit{Astrophys.\  J.} \textbf{128}, 664.
\href{https://doi.org/10.1086/146579}{DOI}.
\href{https://ui.adsabs.harvard.edu/abs/1958ApJ...128..664P/abstract}{ADS}.

\bibitem[\protect\citeauthoryear{{Parker}}{1964}]{P64}
Parker, E.N.: 1964,
Dynamical properties of stellar coronas and stellar winds, I:
Integration of the momentum equation.
\textit{Astrophys.\  J.} \textbf{139}, 72.
\href{https://doi.org/10.1086/147740}{DOI}.
\href{https://ui.adsabs.harvard.edu/abs/1964ApJ...139...72P/abstract}{ADS}.

\bibitem[\protect\citeauthoryear{{Pecora et al.}}{2023}]{Pc23}
Pecora, F., Yang, Y., Gibson, S.E., Viall, N.M., Chhiber, R.,
DeForest, C.E., Matthaeus, W.H.: 2023,
Magnetohydrodynamic turbulence simulations: A testing ground for PUNCH.
\textit{Solar Phys.,} in preparation.

\bibitem[\protect\citeauthoryear{{Pineda and Villadsen}}{2023}]{PV23}
Pineda, J.S., Villadsen, J.: 2023,
Coherent radio bursts from known M-dwarf planet-host YZ Ceti.
\textit{Nature Astron.} \textbf{7}, 569.
\href{https://doi.org/10.1038/s41550-023-01914-0}{DOI}.
\href{https://ui.adsabs.harvard.edu/abs/2023NatAs...7..569P/abstract}{ADS}.

\bibitem[\protect\citeauthoryear{{Pinto et al.}}{2011}]{Pi11}
Pinto, R.F., Brun, A.S., Jouve, L., Grappin, R.: 2011,
Coupling the solar dynamo and the corona: Wind properties, mass, and
momentum losses during an activity cycle.
\textit{Astrophys.\  J.} \textbf{737}, 72.
\href{https://doi.org/10.1088/0004-637X/737/2/72}{DOI}.
\href{https://ui.adsabs.harvard.edu/abs/2011ApJ...737...72P/abstract}{ADS}.

\bibitem[\protect\citeauthoryear{{Pizzo et al.}}{1983}]{Pz83}
Pizzo, V., Schwenn, R., Marsch, E., Rosenbauer, H.,
M\"{u}hlh\"{a}user, K.-H., Neubauer, F.M.: 1983,
Determination of the solar wind angular momentum flux from the
Helios data: An observational text of the Weber and Davis theory.
\textit{Astrophys.\  J.} \textbf{271}, 335.
\href{https://doi.org/10.1086/161200}{DOI}.
\href{https://ui.adsabs.harvard.edu/abs/1983ApJ...271..335P/abstract}{ADS}.

\bibitem[\protect\citeauthoryear{{Pneuman and Kopp}}{1971}]{PK71}
Pneuman, G.W., Kopp, R.A.: 1971,
Gas-magnetic field interactions in the solar corona.
\textit{Solar Phys.} \textbf{18}, 258.
\href{https://doi.org/10.1007/BF00145940}{DOI}.
\href{https://ui.adsabs.harvard.edu/abs/1971SoPh...18..258P/abstract}{ADS}.

\bibitem[\protect\citeauthoryear{{Pulupa et al.}}{2017}]{Pu17}
Pulupa, M., Bale, S.D., Bonnell, J.W., Bowen, T.A., Carruth, N.,
Goetz, K., Gordon, D., Harvey, P.R., Maksimovic, M.,
Mart\'{\i}nez-Oliveros, J.C., Moncuquet, M., Saint-Hilaire, P.,
Seitz, D., Sundkvist, D.: 2017,
The Solar Probe Plus Radio Frequency Spectrometer: Measurement
requirements, analog design, and digital signal processing.
\textit{J.\  Geophys.\  Res.} \textbf{122}, 2836.
\href{https://doi.org/10.1002/2016JA023345}{DOI}.
\href{https://ui.adsabs.harvard.edu/abs/2017JGRA..122.2836P/abstract}{ADS}.

\bibitem[\protect\citeauthoryear{{Raouafi et al.}}{2023}]{Rao23}
Raouafi, N.E., Matteini, L., Squire, J., Badman, S.T., Velli, M.,
Klein, K.G., Chen, C.H.K., Matthaeus, W.H., Szabo, A., Linton, M.,
Allen, R.C., Szalay, J.R., Bruno, R., Decker, R.B., Akhavan-Tafti, M.,
Agapitov, O.V., Bale, S.D., Banyopadhyay, R., Battams, K.,
Ber\v{c}i\v{c}, L., Bourouaine, S., Bowen, T.A., Cattell, C.,
Chandran, B.D.G., Chhiber, R., Cohen, C.M.S., D'Amicis, R.,
Giacalone, J., Hess, P., Howard, R.A., Horbury, T.S.,
Jagarlamudi, V.K., Joyce, C.J., Kasper, J.C., Kinnison, J., Laker, R.,
Liewer, P., Malaspina, D.M., Mann, I., McComas, D.J.,
Niembro-Hernandez, T., Nieves-Chinchilla, T., Panasenco, O.,
Pokorn\'{y}, P., Pusack, A., Pulupa, M., Perez, J.C., Riley, P.,
Rouillard, A.P., Shi, C., Stenborg, G., Tenerani, A., Verniero, J.L.,
Viall, N., Vourlidas, A., Wood, B.E., Woodham, L.D., Woolley, T.: 2023,
Parker Solar Probe: Four years of discoveries at solar cycle minimum.
\textit{Space Sci.\  Rev.} \textbf{219}, 8.
\href{https://doi.org/10.1007/s11214-023-00952-4}{DOI}.
\href{https://ui.adsabs.harvard.edu/abs/2023SSRv..219....8R/abstract}{ADS}.

\bibitem[\protect\citeauthoryear{{Reiss et al.}}{2019}]{Rs19}
Reiss, M.A., MacNeice, P.J., Mays, L.M., Arge, C.N., M\"{o}stl, C.,
Nikolic, L., Amerstorfer, T.: 2019,
Forecasting the ambient solar wind with numerical models, I: On the
implementation of an operational framework.
\textit{Astrophys.\  J.\  Suppl.} \textbf{240}, 35.
\href{https://doi.org/10.3847/1538-4365/aaf8b3}{DOI}.
\href{https://ui.adsabs.harvard.edu/abs/2019ApJS..240...35R/abstract}{ADS}.

\bibitem[\protect\citeauthoryear{{R\'{e}ville and Brun}}{2017}]{RB17}
R\'{e}ville, V., Brun, A.S.: 2017,
Global solar magnetic field organization in the outer corona:
Influence on the solar wind speed and mass flux over the cycle.
\textit{Astrophys.\  J.} \textbf{850}, 45.
\href{https://doi.org/10.3847/1538-4357/aa9218}{DOI}.
\href{https://ui.adsabs.harvard.edu/abs/2017ApJ...850...45R/abstract}{ADS}.

\bibitem[\protect\citeauthoryear{{Riley et al.}}{2006}]{Ri06}
Riley, P., Linker, J.A., Miki\'{c}, Z., Lionello, R., Ledvina, S.A.,
Luhmann, J.G.: 2006,
A comparison between global solar magnetohydrodynamic and potential
field source surface model results.
\textit{Astrophys.\  J.} \textbf{653}, 1510.
\href{https://doi.org/10.1086/508565}{DOI}.
\href{https://ui.adsabs.harvard.edu/abs/2006ApJ...653.1510R/abstract}{ADS}.

\bibitem[\protect\citeauthoryear{{Riley and Lionello}}{2011}]{RL11}
Riley, P., Lionello, R.: 2011,
Mapping solar wind streams from the Sun to 1 AU: A comparison of techniques.
\textit{Solar Phys.} \textbf{270}, 575.
\href{https://doi.org/10.1007/s11207-011-9766-x}{DOI}.
\href{https://ui.adsabs.harvard.edu/abs/2011SoPh..270..575R/abstract}{ADS}.

\bibitem[\protect\citeauthoryear{{Ruffolo et al.}}{2020}]{Ru20}
Ruffolo, D., Matthaeus, W.H., Chhiber, R., Usmanov, A.V., Yang, Y.,
Bandyopadhyay, R., Parashar, T.N., Goldstein, M.L., DeForest, C.E.,
Wan, M., Chasapis, A., Maruca, B.A., Velli, M., Kasper, J.C.: 2020,
Shear-driven transition to isotropically turbulent solar wind outside
the Alfv\'{e}n critical zone.
\textit{Astrophys.\  J.} \textbf{902}, 94.
\href{https://doi.org/10.3847/1538-4357/abb594}{DOI}.
\href{https://ui.adsabs.harvard.edu/abs/2020ApJ...902...94R/abstract}{ADS}.

\bibitem[\protect\citeauthoryear{{Sakurai}}{1985}]{Sk85}
Sakurai, T.: 1985,
Magnetic stellar winds: A 2-D generalization of the Weber-Davis model.
\textit{Astron.\  Astrophys.} \textbf{152}, 121.
\href{https://ui.adsabs.harvard.edu/abs/1985A&A...152..121S/abstract}{ADS}.

\bibitem[\protect\citeauthoryear{{Sakurai}}{1990}]{Sk90}
Sakurai, T.: 1990,
Magnetohydrodynamic solar/stellar wind models.
\textit{Comput.\  Phys.\  Rep.} \textbf{12}, 247.
\href{https://doi.org/10.1016/0167-7977(90)90013-V}{DOI}.
\href{https://ui.adsabs.harvard.edu/abs/1990CoPhR..12..247S/abstract}{ADS}.

\bibitem[\protect\citeauthoryear{{Sanchez-Diaz et al.}}{2017}]{SD17}
Sanchez-Diaz, E., Rouillard, A.P., Davies, J.A., Lavraud, B., Sheeley, N.R.,
Pinto, R.F., Kilpua, E., Plotnikov, I., Genot, V.: 2017,
Observational evidence for the associated formation of blobs and
raining inflows in the solar corona.
\textit{Astrophys.\  J.\  Lett.} \textbf{835}, L7.
\href{https://doi.org/10.3847/2041-8213/835/1/L7}{DOI}.
\href{https://ui.adsabs.harvard.edu/abs/2017ApJ...835L...7S/abstract}{ADS}.

\bibitem[\protect\citeauthoryear{{Saur et al.}}{2013}]{Saur13}
Saur, J., Grambusch, T., Duling, S., Neubauer, F.M., Simon, S.: 2013,
Magnetic energy fluxes in sub-Alfv\'{e}nic planet star and moon planet
interactions.
\textit{Astron.\  Astrophys.} \textbf{552}, A119.
\href{https://doi.org/10.1051/0004-6361/201118179}{DOI}.
\href{https://ui.adsabs.harvard.edu/abs/2013A&A...552A.119S/abstract}{ADS}.

\bibitem[\protect\citeauthoryear{{Savage et al.}}{2012}]{Sv12}
Savage, S.L., McKenzie, D.E., Reeves, K.K.: 2012,
Re-interpretation of supra-arcade downflows in solar flares.
\textit{Astrophys.\  J.\  Lett.} \textbf{747}, L40.
\href{https://doi.org/10.1088/2041-8205/747/2/L40}{DOI}.
\href{https://ui.adsabs.harvard.edu/abs/2012ApJ...747L..40S/abstract}{ADS}.

\bibitem[\protect\citeauthoryear{{Schatten et al.}}{1969}]{Sc69}
Schatten, K.H., Wilcox, J.M., Ness, N.F.: 1969,
A model of interplanetary and coronal magnetic fields.
\textit{Solar Phys.} \textbf{6}, 442.
\href{https://doi.org/10.1007/BF00146478}{DOI}.
\href{https://ui.adsabs.harvard.edu/abs/1969SoPh....6..442S/abstract}{ADS}.

\bibitem[\protect\citeauthoryear{{Scherrer et al.}}{1995}]{Sc95}
Scherrer, P.H., Bogart, R.S., Bush, R.I., Hoeksema, J.T., Kosovichev, A.G.,
Schou, J., Rosenberg, W., Springer, L., Tarbell, T.D., Title, A.,
Wolfson, C.J., Zayer, I., MDI Engineering Team: 1995,
The Solar Oscillations Investigation: Michelson Doppler Imager.
\textit{Solar Phys.} \textbf{162}, 129.
\href{https://doi.org/10.1007/BF00733429}{DOI}.
\href{https://ui.adsabs.harvard.edu/abs/1995SoPh..162..129S/abstract}{ADS}.

\bibitem[\protect\citeauthoryear{{Schwadron et al.}}{2010}]{Sw10}
Schwadron, N.A., Connick, D.E., Smith, C.: 2010,
Magnetic flux balance in the heliosphere.
\textit{Astrophys.\  J.\  Lett.} \textbf{722}, L132.
\href{https://doi.org/10.1088/2041-8205/722/2/L132}{DOI}.
\href{https://ui.adsabs.harvard.edu/abs/2010ApJ...722L.132S/abstract}{ADS}.

\bibitem[\protect\citeauthoryear{{Sheeley and Wang}}{2014}]{SW14}
Sheeley, N.R., Wang, Y.-M.: 2014,
Coronal inflows during the interval 1996--2014.
\textit{Astrophys.\  J.} \textbf{797}, 10.
\href{https://doi.org/10.1088/0004-637X/797/1/10}{DOI}.
\href{https://ui.adsabs.harvard.edu/abs/2014ApJ...797...10S/abstract}{ADS}.

\bibitem[\protect\citeauthoryear{{Sheeley et al.}}{1997}]{Sh97}
Sheeley, N.R., Wang, Y.-M., Hawley, S.H., Brueckner, G.E., Dere, K.P.,
Howard, R.A., Koomen, M.J., Korendyke, C.M., Michels, D.J., Paswaters, S.E.,
Socker, D.G., St.~Cyr, O.C., Wang, D., Lamy, P.L., Llebaria, A., Schwenn, R.,
Simnett, G.M., Plunkett, S., Biesecker, D.A.: 1997,
Measurements of flow speeds in the corona between 2 and 30 $R_{\odot}$. 
\textit{Astrophys.\  J.} \textbf{484}, 472.
\href{https://doi.org/10.1086/304338}{DOI}.
\href{https://ui.adsabs.harvard.edu/abs/1997ApJ...484..472S/abstract}{ADS}.

\bibitem[\protect\citeauthoryear{{Shkolnik et al.}}{2008}]{Shk08}
Shkolnik, E., Bohlender, D.A., Walker, G.A.H., Collier Cameron, A.: 2008,
The on-off nature of star-planet interactions.
\textit{Astrophys.\  J.} \textbf{676}, 628.
\href{https://doi.org/10.1086/527351}{DOI}.
\href{https://ui.adsabs.harvard.edu/abs/2008ApJ...676..628S/abstract}{ADS}.

\bibitem[\protect\citeauthoryear{{Shodhan et al.}}{2000}]{Sh00}
Shodhan, S., Crooker, N.U., Kahler, S.W., Fitzenreiter, R.J., Larson, D.E.,
Lepping, R.P., Siscoe, G.L., Gosling, J.T.: 2000,
Counterstreaming electrons in magnetic clouds.
\textit{J.\  Geophys.\  Res.} \textbf{105}, 27261.
\href{https://doi.org/10.1029/2000JA000060}{DOI}.
\href{https://ui.adsabs.harvard.edu/abs/2000JGR...10527261S/abstract}{ADS}.

\bibitem[\protect\citeauthoryear{{Skumanich}}{1972}]{Sk72}
Skumanich, A.: 1972,
Time scales for Ca II emission decay, rotational braking, and
lithium depletion.
\textit{Astrophys.\  J.} \textbf{171}, 565.
\href{https://doi.org/10.1086/151310}{DOI}.
\href{https://ui.adsabs.harvard.edu/abs/1972ApJ...171..565S/abstract}{ADS}.

\bibitem[\protect\citeauthoryear{{Smith et al.}}{2013}]{Sm13}
Smith, C.W., Schwadron, N.A., DeForest, C.E.: 2013,
Decline and recovery of the interplanetary magnetic field during
the protracted solar minimum.
\textit{Astrophys.\  J.} \textbf{775}, 59.
\href{https://doi.org/10.1088/0004-637X/775/1/59}{DOI}.
\href{https://ui.adsabs.harvard.edu/abs/2013ApJ...775...59S/abstract}{ADS}.

\bibitem[\protect\citeauthoryear{{Sturrock and Hartle}}{1966}]{SH66}
Sturrock, P.A., Hartle, R.E.: 1966,
Two-fluid model of the solar wind.
\textit{Phys.\  Rev.\  Lett.} \textbf{16}, 628.
\href{https://doi.org/10.1103.PhysRevLett.16.628}{DOI}.
\href{https://ui.adsabs.harvard.edu/abs/1966PhRvL..16..628S/abstract}{ADS}.

\bibitem[\protect\citeauthoryear{{Suess}}{1982}]{Su82}
Suess, S.T.: 1982,
Unsteady, thermally conductive coronal flow.
\textit{Astrophys.\  J.} \textbf{259}, 880.
\href{https://doi.org/10.1086/160222}{DOI}.
\href{https://ui.adsabs.harvard.edu/abs/1982ApJ...259..880S/abstract}{ADS}.

\bibitem[\protect\citeauthoryear{{Suess et al.}}{1999}]{Su99}
Suess, S.T., Wang, A.-H., Wu, S.T., Poletto, G., McComas, D.J.: 1999,
A two-fluid, MHD coronal model.
\textit{J.\  Geophys.\  Res.} \textbf{104}, 4697.
\href{https://doi.org/10.1029/1998JA900086}{DOI}.
\href{https://ui.adsabs.harvard.edu/abs/1999JGR...104.4697S/abstract}{ADS}.

\bibitem[\protect\citeauthoryear{{Tasnim and Cairns}}{2016}]{TC16}
Tasnim, S., Cairns, I.H.: 2016,
An equatorial solar wind model with angular momentum conservation and
nonradial magnetic fields and flow velocities at an inner boundary.
\textit{J.\  Geophys.\  Res.} \textbf{121}, 4966.
\href{https://doi.org/10.1002/2016JA022725}{DOI}.
\href{https://ui.adsabs.harvard.edu/abs/2016JGRA..121.4966T/abstract}{ADS}.

\bibitem[\protect\citeauthoryear{{Tasnim et al.}}{2019}]{TC19}
Tasnim, S., Cairns, I.H., Li, B., Wheatland, M.S.: 2019,
Mapping magnetic field lines for an accelerating solar wind.
\textit{Solar Phys.} \textbf{294}, 155.
\href{https://doi.org/10.1007/s11207-019-1541-4}{DOI}.
\href{https://ui.adsabs.harvard.edu/abs/2019SoPh..294..155T/abstract}{ADS}.

\bibitem[\protect\citeauthoryear{{Tasnim et al.}}{2018}]{TC18}
Tasnim, S., Cairns, I.H., Wheatland, M.S.: 2018,
A generalized equatorial model for the accelerating solar wind.
\textit{J.\  Geophys.\  Res.} \textbf{123}, 1061.
\href{https://doi.org/10.1002/2017JA024532}{DOI}.
\href{https://ui.adsabs.harvard.edu/abs/2018JGRA..123.1061T/abstract}{ADS}.

\bibitem[\protect\citeauthoryear{{Taylor}}{1938}]{T38}
Taylor, G.I.: 1938,
The spectrum of turbulence.
\textit{Proc.\  Roy.\  Soc.\  London A} \textbf{164}, 476.
\href{https://doi.org/10.1098/rspa.1938.0032}{DOI}.
\href{https://ui.adsabs.harvard.edu/abs/1938RSPSA.164..476T/abstract}{ADS}.

\bibitem[\protect\citeauthoryear{{Telloni et al.}}{2021}]{Td21}
Telloni, D., Andretta, V., Antonucci, E., Bemporad, A., Capuano, G.E.,
Fineschi, S., Giordano, S., Habbal, S., Perrone, D., Pinto, R.F.,
Sorriso-Valvo, L., Spadaro, D., Susino, R., Woodham, L.D., Zank, G.P.,
Romoli, M., Bale, S.D., Kasper, J.C., Auch\`{e}re, F., Bruno, R.,
Capobianco, G., Case, A.W., Casini, C., Casti, M., Chioetto, P.,
Corso, A.J., Da Deppo, V., De Leo, Y., Dudok de Wit, T., Frassati, F.,
Frassetto, F., Goetz, K., Guglielmino, S.L., Harvey, P.R., Heinzel, P.,
Jerse, G., Korreck, K.E., Landini, F., Larson, D., Liberatore, A.,
Livi, R., MacDowall, R.J., Magli, E., Malaspina, D.M., Massone, G.,
Messerotti, M., Moses, J.D., Naletto, G., Nicolini, G., Nistic\`{o}, G.,
Panasenco, O., Pancrazzi, M., Pelizzo, M.G., Pulupa, M., Reale, F.,
Romano, P., Sasso, C., Sch\"{u}hle, U., Stangalini, M., Stevens, M.L.,
Strachan, L., Straus, T., Teriaca, L., Uslenghi, M., Velli, M.,
Verscharen, D., Volpicelli, C.A., Whittlesey, P., Zangrilli, L.,
Zimbardo, G., Zuppella, P.: 2021,
Exploring the solar wind from its source on the corona into the inner
heliosphere during the first Solar Orbiter--Parker Solar Probe quadrature.
\textit{Astrophys.\  J.\  Lett.} \textbf{920}, L14.
\href{https://doi.org/10.3847/2041-8213/ac282f}{DOI}.
\href{https://ui.adsabs.harvard.edu/abs/2021ApJ...920L..14T/abstract}{ADS}.

\bibitem[\protect\citeauthoryear{{Tenerani et al.}}{2016}]{Te16}
Tenerani, A., Velli, M., DeForest, C.E.: 2016,
Inward motions in the outer solar corona between 7 and 12 $R_{\odot}$:
Evidence for waves or magnetic reconnection jets?
\textit{Astrophys.\  J.\  Lett.} \textbf{825}, L3.
\href{https://doi.org/10.3847/2041-8205/825/1/L3}{DOI}.
\href{https://ui.adsabs.harvard.edu/abs/2016ApJ...825L...3T/abstract}{ADS}.

\bibitem[\protect\citeauthoryear{{Townsend et al.}}{2010}]{Tw10}
Townsend, R.H.D., Oksala, M.E., Cohen, D.H., Owocki, S.P.,
ud-Doula, A.: 2010,
Discovery of rotational braking in the magnetic helium-strong star
sigma Orionis E.
\textit{Astrophys.\  J.\  Lett.} \textbf{714}, L318.
\href{https://doi.org/10.1088/2041-8205/714/2/L318}{DOI}.
\href{https://ui.adsabs.harvard.edu/abs/2010ApJ...714L.318T/abstract}{ADS}.

\bibitem[\protect\citeauthoryear{{ud-Doula}}{2017}]{uD17}
ud-Doula, A.: 2017,
Magnetic fields in massive stars and magnetically confined winds.
\textit{Astron.\  Nachrichten} \textbf{338}, 944.
\href{https://doi.org/10.1002/asna.201713394}{DOI}.
\href{https://ui.adsabs.harvard.edu/abs/2017AN....338..944U/abstract}{ADS}.

\bibitem[\protect\citeauthoryear{{Usmanov et al.}}{2018}]{Us18}
Usmanov, A.V., Matthaeus, W.H., Goldstein, M.L., Chhiber, R.: 2018,
The steady global corona and solar wind: A three-dimensional MHD
simulation with turbulence transport and heating.
\textit{Astrophys.\  J.} \textbf{865}, 25.
\href{https://doi.org/10.3847/1538-4357/aad687}{DOI}.
\href{https://ui.adsabs.harvard.edu/abs/2018ApJ...865...25U/abstract}{ADS}.

\bibitem[\protect\citeauthoryear{{V\'{a}squez et al.}}{2003}]{Va03}
V\'{a}squez, A.M., van Ballegooijen, A.A., Raymond, J.C.: 2003,
The effect of proton temperature anisotropy on the solar minimum
corona and wind.
\textit{Astrophys.\  J.} \textbf{598}, 1361.
\href{https://doi.org/10.1086/379008}{DOI}.
\href{https://ui.adsabs.harvard.edu/abs/2003ApJ...598.1361V/abstract}{ADS}.

\bibitem[\protect\citeauthoryear{{Velli}}{1993}]{Ve93}
Velli, M.: 1993,
On the propagation of ideal, linear Alfv\'{e}n waves in radially
stratified stellar atmospheres and winds.
\textit{Astron.\  Astrophys.} \textbf{270}, 304.
\href{https://ui.adsabs.harvard.edu/abs/1993A&A...270..304V/abstract}{ADS}.

\bibitem[\protect\citeauthoryear{{Velli}}{1994}]{Ve94}
Velli, M.: 1994,
From supersonic winds to accretion: Comments on the stability of stellar
winds and related flows.
\textit{Astrophys.\  J.\  Lett.} \textbf{432}, L55.
\href{https://doi.org/10.1086/187510}{DOI}.
\href{https://ui.adsabs.harvard.edu/abs/1994ApJ...432L..55V/abstract}{ADS}.

\bibitem[\protect\citeauthoryear{{Velli}}{2001}]{Ve01}
Velli, M.: 2001,
Hydrodynamics of the solar wind expansion.
\textit{Astrophys.\  Space Sci.} \textbf{277}, 157.
\href{https://doi.org/10.1023/A:1012237708634}{DOI}
\href{https://ui.adsabs.harvard.edu/abs/2001Ap&SS.277..157V/abstract}{ADS}.

\bibitem[\protect\citeauthoryear{{Verscharen et al.}}{2021a}]{Vs21a}
Verscharen, D., Bale, S.D., Velli, M.: 2021a,
Flux conservation, radial scalings, Mach numbers, and critical distances
in the solar wind: Magnetohydrodynamics and Ulysses observations.
\textit{Mon.\  Not.\  Roy.\  Astron.\  Soc.} \textbf{506}, 4993.
\href{https://doi.org/10.1093/mnras/stab2051}{DOI}.
\href{https://ui.adsabs.harvard.edu/abs/2021MNRAS.506.4993V/abstract}{ADS}.

\bibitem[\protect\citeauthoryear{{Verscharen et al.}}{2021b}]{Vs21b}
Verscharen, D., Stansby, D., Finley, A.J., Owen, C.J., Horbury, T.,
Maksimovic, M., Velli, M., Bale, S.D., Louarn, P., Fedorov, A.,
Bruno, R., Livi, S., Khotyaintsev, Y.V., Vecchio, A., Lewis, G.R.,
Anekallu, C., Kelly, C.W., Watson, G., Kataria, D.O., O'Brien, H.,
Evans, V., Angelini, V., Solar Orbiter SWA, MAG, and RPW Teams: 2021b,
The angular-momentum flux in the solar wind observed during
Solar Orbiter's first orbit.
\textit{Astron.\  Astrophys.} \textbf{656}, A28.
\href{https://doi.org/10.1051/0004-6361/202140956}{DOI}.
\href{https://ui.adsabs.harvard.edu/abs/2021A&A...656A..28V/abstract}{ADS}.

\bibitem[\protect\citeauthoryear{{Viall et al.}}{2021}]{Vn21}
Viall, N.M., DeForest, C.E., Kepko, L.: 2021,
Mesoscale structure in the solar wind.
\textit{Front.\  Astron.\  Space Sci.} \textbf{8}, 139.
\href{https://doi.org/10.3389/fspas.2021.735034}{DOI}.
\href{https://ui.adsabs.harvard.edu/abs/2021FrASS...8..139V/abstract}{ADS}.

\bibitem[\protect\citeauthoryear{{Vidotto}}{2021}]{Vi21}
Vidotto, A.A.: 2021,
The evolution of the solar wind.
\textit{Liv.\  Rev.\  Solar Phys.} \textbf{18}, 3.
\href{https://doi.org/10.1007/s41116-021-00029-w}{DOI}.
\href{https://ui.adsabs.harvard.edu/abs/2021LRSP...18....3V/abstract}{ADS}.

\bibitem[\protect\citeauthoryear{{Weber and Davis}}{1967}]{WD67}
Weber, E.J., Davis, L., Jr.: 1967,
The angular momentum of the solar wind.
\textit{Astrophys.\  J.} \textbf{148}, 217.
\href{https://doi.org/10.1086/149138}{DOI}.
\href{https://ui.adsabs.harvard.edu/abs/1967ApJ...148..217W/abstract}{ADS}.

\bibitem[\protect\citeauthoryear{{Wexler et al.}}{2021}]{Wx21}
Wexler, D.B., Stevens, M.L., Case, A.W., Song. P.: 2021,
Alfv\'{e}n speed transition zone in the solar corona.
\textit{Astrophys.\  J.\  Lett.} \textbf{919}, L33.
\href{https://doi.org/10.3847/2041-8213/ac25fa}{DOI}.
\href{https://ui.adsabs.harvard.edu/abs/2021ApJ...919L..33W/abstract}{ADS}.

\bibitem[\protect\citeauthoryear{{Xu and Borovsky}}{2015}]{XB15}
Xu, F., Borovsky, J.E.: 2015,
A new four-plasma categorization scheme for the solar wind.
\textit{J.\  Geophys.\  Res.} \textbf{120}, 70.
\href{https://doi.org/10.1002/2014JA020412}{DOI}.
\href{https://ui.adsabs.harvard.edu/abs/2015JGRA..120...70X/abstract}{ADS}.

\bibitem[\protect\citeauthoryear{{Zank et al.}}{2022}]{Zn22}
Zank, G.P., Zhao, L.-L., Adhikari, L., Telloni, D., Kasper, J.C.,
Stevens, M., Rahmati, A., Bale, S.D.: 2022,
Turbulence in the sub-Alfv\'{e}nic solar wind.
\textit{Astrophys.\  J.\  Lett.} \textbf{926}, L16.
\href{https://doi.org/10.3847/2041-8213/ac51da}{DOI}.
\href{https://ui.adsabs.harvard.edu/abs/2022ApJ...926L..16Z/abstract}{ADS}.

\bibitem[\protect\citeauthoryear{{Zhang et al.}}{2022}]{Zg22}
Zhang, J., Huang, S.Y., Yuan, Z.G., Jiang, K., Xu, S.B.,
Bandyopadhyay, R., Wei, Y.Y., Xiong, Q.Y., Wang, Z., Yu, L.,
Lin, R.T.: 2022,
Higher-order turbulence statistics in sub-Alfv\'{e}nic solar wind
observed by Parker Solar Probe.
\textit{Astrophys.\  J.} \textbf{937}, 70.
\href{https://doi.org/10.3847/1538-4357/ac8c34}{DOI}.
\href{https://ui.adsabs.harvard.edu/abs/2022ApJ...937...70Z/abstract}{ADS}.

\bibitem[\protect\citeauthoryear{{Zhao et al.}}{2022}]{Zh22}
Zhao, L.-L., Zank, G.P., Telloni, D., Stevens, M., Kasper, J.C.,
Bale, S.D.: 2022,
The turbulent properties of the sub-Alfv\'{e}nic solar wind
measured by the Parker Solar Probe.
\textit{Astrophys.\  J.\  Lett.} \textbf{928}, L15.
\href{https://doi.org/10.3847/2041-8213/ac5fb0}{DOI}.
\href{https://ui.adsabs.harvard.edu/abs/2022ApJ...928L..15Z/abstract}{ADS}.

\bibitem[\protect\citeauthoryear{{Zhao and Hoeksema}}{2010}]{ZH10}
Zhao, X.P., Hoeksema, J.T.: 2010,
The magnetic field at the inner boundary of the heliosphere around
solar minimum.
\textit{Solar Phys.} \textbf{266}, 379.
\href{https://doi.org/10.1007/s11207-010-9618-0}{DOI}.
\href{https://ui.adsabs.harvard.edu/abs/2010SoPh..266..379Z/abstract}{ADS}.

\end{thebibliography}
\end{document}